\documentclass[sigconf]{acmart}

\AtBeginDocument{%
  }

\usepackage{algorithm}
\usepackage[noend]{algorithmic}
\usepackage{graphicx}
\graphicspath{{figs/}}
\usepackage{multirow}
\usepackage{makecell}
\usepackage{xspace}
\usepackage{subcaption}
\usepackage{colortbl}
\usepackage{amsmath}

\allowdisplaybreaks[4]
\usepackage{amsthm}
\usepackage{enumitem}
\usepackage[table]{xcolor}

\copyrightyear{2026}
\acmYear{2026}
\setcopyright{cc}
\setcctype{by-nc-nd}
\acmConference[ICPP '26]{Proceedings of the 55th International Conference on Parallel Processing}{September 28-October 01, 2026}{Singapore, Singapore}
\acmBooktitle{Proceedings of the 55th International Conference on Parallel Processing (ICPP '26), September 28-October 01, 2026, Singapore, Singapore}
\acmDOI{10.1145/3832810.3832835}
\acmISBN{979-8-4007-2657-6/2026/09}

\begin{document}

\title{Scaling Synthetic-Image Pre-Training for Federated Fine-Tuning of Large Vision Models}



\author{Qianpiao Ma}
\affiliation{%
  \institution{Nanjing University of Science and Technology}
  \city{Nanjing}
  \state{Jiangsu}
  \country{China}}
\email{maqianpiao@njust.edu.cn}

\author{Xiaozhu Song}
\affiliation{%
  \institution{Nanjing University of Science and Technology}
  \city{Nanjing}
  \state{Jiangsu}
  \country{China}}
\email{songxiaozhu@njust.edu.cn}

\author{Junlong Zhou}
\authornote{Corresponding author.}
\affiliation{%
  \institution{Nanjing University of Science and Technology}
  \city{Nanjing}
  \state{Jiangsu}
  \country{China}}
\email{jlzhou@njust.edu.cn}

\author{Yue Zeng}
\affiliation{%
  \institution{Nanjing University of Science and Technology}
  \city{Nanjing}
  \state{Jiangsu}
  \country{China}}
\email{zengyue@njust.edu.cn}

\author{Jianchun Liu}
\affiliation{%
 \institution{University of Science and Technology of China}
 \city{Hefei}
 \state{Anhui}
 \country{China}}
\email{jcliu17@ustc.edu.cn}

\author{Huaqing Tu}
\affiliation{%
 \institution{Zhejiang Gongshang University}
 \city{Hangzhou}
 \state{Zhejiang}
 \country{China}}
\email{thq@mail.zjgsu.edu.cn}

\renewcommand{\shortauthors}{Ma et al.}
\renewcommand{\algorithmicrequire}{\textbf{Input:}}
\renewcommand{\algorithmicensure}{\textbf{Output:}}
\def\ie{\textit{i.e.}\xspace}
\def\etal{\textit{et al.}\xspace}
\def\etc{\textit{etc.}\xspace}
\def\whp{\textit{w.h.p.}\xspace}
\def\eg{\textit{e.g.}\xspace}
\def\iid{\textit{i.i.d.}\xspace}
\def\st{\xspace\textbf{s.t.}\xspace}
\theoremstyle{colon}
\renewcommand{\figurename}{Fig.}

\newtheorem{Theorem}{Theorem}
\newtheorem{axiom}[theorem]{Axiom}
\newtheorem{mydefinition}{Definition}
\newtheorem{mylemma}{Lemma}
\newtheorem{remark}{Remark}
\newtheorem{protocol}{Protocol}
\newtheorem{assumption}{Assumption}


\begin{abstract}

Federated fine-tuning (FedFT) enables adapting pre-trained large vision models (LVMs) on distributed, privacy-sensitive devices, while its practical deployment is hindered by three critical challenges: resource constraints, system heterogeneity, and non-IID data. While prior studies partially address these issues, \eg, by pre-training initial models on synthetic images to mitigate the adverse effects of non-IID data, or leveraging parameter-efficient fine-tuning (PEFT) methods like low-rank adaptation (LoRA) to reduce resource consumption, they remain inadequate and fragmented. Specifically, existing synthetic image generation methods fail to capture device-specific feature distributions, while current PEFT-based FedFT methods often undervalue weaker devices that may provide critical information. More importantly, the separate optimization of pre-training and FedFT neglects their inherent connection, lacking a holistic perspective to maximize training efficiency. To overcome these limitations, we propose FeDiSyn, a unified framework that holistically considers the interplay between pre-training and FedFT to minimize the overall LVM training time. Specifically, FeDiSyn introduces: (i) a scaling law for FedFT pre-training to determine the optimal number of synthetic images, balancing pre-training benefit against generation/pre-training cost, (ii) diffusion-based synthetic image generation that captures device-specific feature distributions for pre-training to tackle non-IID data, and (iii) a contribution-aware LoRA configuration and bandwidth allocation algorithm for FedFT to ensure that informative devices are effectively utilized while addressing system heterogeneity. Experimental results on the real-world testbed demonstrate that FeDiSyn reduces completion time by over 52.5\% and communication cost by over 97.2\%, while achieving comparable accuracy to state-of-the-art solutions.

\end{abstract}

\ccsdesc[500]{Computing methodologies~Distributed artificial intelligence}
\ccsdesc[500]{Human-centered computing~Mobile computing}

\keywords{Federated fine-tuning, Synthetic image, Pre-training, Scaling law, Large vision models}


\maketitle

\section{Introduction}\label{sec:introduction}

Since the landmark success of AlexNet in 2012 \cite{krizhevsky2012imagenet}, deep learning powered by large-scale labeled datasets has revolutionized the field of image classification. Initially, most vision models are trained via supervised learning on relatively small datasets. More recently, however, a two-stage paradigm has emerged for training large vision models (LVM): first pre-training on massive datasets, then fine-tuning on downstream task-specific data. In particular, pure transformer-based architectures such as the vision transformer (ViT) \cite{dosovitskiy2020image} have demonstrated more outstanding performance than the convolutional networks when pre-trained at scale.

In parallel, the Internet of Things (IoT) \cite{zhu2020toward}\cite{zhou2022swarm} has grown explosively, generating massive amounts of data from edge devices that offers a rich source for training LVMs. However, centralized training on large-scale datasets is increasingly limited by privacy regulations (\eg, GDPR) and by bandwidth bottlenecks when accessing local data. This motivates the adoption of federated fine-tuning (FedFT) \cite{zhang2023fedpetuning}, a technique that integrates model fine-tuning into the federated learning (FL) framework \cite{mcmahan2017communication}. In a typical FedFT setting, a parameter server first broadcasts a pre-trained global model to a set of distributed devices. Each device fine-tunes the model on its local data and sends the updated parameters back to the server for aggregation. This cycle of local fine-tuning and global aggregation repeats until the model converges or reaches a target accuracy.

\begin{table*}[t]\centering
\small{
\setlength{\tabcolsep}{8pt}
\caption{Comparison of FedFT methods for LVMs.}\label{tbl:FLcomparison}
\vspace{-3mm}
\begin{tabular}{c||c|c|c|c|c|c} 
\hline
\multicolumn{1}{c||}{\makecell{Method}} & \makecell{Synthetic Image \\for Pre-training} & \makecell{PEFT\\Method} & \makecell{Handling Resource \\ Constraints} & \makecell{Handling System\\Heterogeneity} & \makecell{Handling Label \\ Distribution Skew} & \makecell{Handling Feature \\ Distribution Skew}   \\ \hline
PerAda \cite{xie2024perada} & No & Adapter & \checkmark & \texttimes & \texttimes & \texttimes  \\
pFedSeq \cite{peng2025look} & No & Adapter & \checkmark & \texttimes & \texttimes & \texttimes  \\
CaFPT \cite{guo2024explore} & No & Prompt & \checkmark & \texttimes & \texttimes & \texttimes  \\
FedPrefix \cite{sun2023fedperfix} & No & Prompt & \checkmark & \texttimes & \texttimes & \texttimes  \\
SFLF \cite{wang2025federated} & No & LoRA & \checkmark & \texttimes & \texttimes & \texttimes  \\
FedRA \cite{su2024fedra} & No & LoRA & \checkmark & \checkmark & \texttimes & \texttimes  \\
CAFF\cite{pfeiffer2024efficient} & No & LoRA & \checkmark & \checkmark & \texttimes & \texttimes
\\\hline
FPS \cite{chen2023on} & Yes & - & \texttimes & \texttimes & \checkmark & \texttimes  \\
GPT-FL \cite{zhang2025gpt} & Yes & - & \texttimes & \texttimes & \checkmark & \texttimes  \\
FGL \cite{zhang2023federated} & Yes & - & \texttimes & \texttimes & \checkmark & \texttimes
\\\hline
\textbf{FeDiSyn (Ours)} & Yes & LoRA & \checkmark & \checkmark & \checkmark & \checkmark  \\ \hline
\end{tabular}}
\vspace{-2mm}
\end{table*}

\textbf{Challenges of FedFT for LVMs.} Although FedFT offers a promising solution for fine-tuning LVMs leveraging distributed data, it still faces several practical challenges: (1) \textit{Resource Constraints.} Modern LVMs like ViT and SwinT \cite{liu2021Swin}, usually contain billions of parameters, requiring significant computation/communication resources for fine-tuning/transmission. For example, ViT-L/16 requires roughly 3.1GB of traffic per communication and over 2,900 TFLOPs of computation. In contrast, device resources are typically limited \cite{wang2019adaptive}. For example, devices like the NVIDIA Jetson TX2 offer less than 2TFLOPs of computation capacity, while network bandwidth often falls below 100Mbps, rendering full-parameter FedFT infeasible on devices. (2) \textit{System Heterogeneity.} Devices differ widely in hardware capabilities and network conditions. For example, the computational power of an iPhone 16 amounts to only about 12\% of that of an RTX 3090 desktop GPU \cite{li2026fedquad}, while network condition varies significantly, ranging from 4G ($\sim$ 100Mbps), 5G ($\sim$ 500Mbps) to fiber connections ($\sim$ 1Gbps).
As a result, the time required for fine-tuning and transmitting parameters can differ substantially across devices. Faster devices may idle while waiting for slower ones, leading to a \textit{straggler problem} \cite{ma2021fedsa} that reduces overall FedFT efficiency. (3) \textit{Non-IID Data.} Local data on devices are often non-independent and non-identically distributed \cite{ma2025asynchronous}, which involves two aspects. First, the \textit{label distribution} varies across devices, \eg, some classes may be overrepresented or missing entirely on certain devices. Second, even for the same label, the \textit{feature distributions} may differ due to variations in style or domain, \eg, cartoon vs. natural images \cite{yan2025simple}.

\textbf{Status Quo and Limitations.} To address resource constraints in FedFT, some studies \cite{xie2024perada,peng2025look,guo2024explore,sun2023fedperfix,wang2025federated,su2024fedra,pfeiffer2024efficient} have deployed the parameter-efficient fine-tuning (PEFT) methods (\eg, Adapter tuning, LoRA and Prompt tuning). PEFT inserts lightweight trainable modules into the LVM while keeping the pre-trained model frozen. Since these modules typically account for less than 5\% of the pre-trained model's parameters, PEFT significantly reduces the computation and communication overhead. However, most of these works \cite{xie2024perada,peng2025look,guo2024explore,sun2023fedperfix,wang2025federated} simply transplant centralized PEFT methods into distributed settings without accounting for system heterogeneity across devices. A few studies \cite{su2024fedra,pfeiffer2024efficient} attempt to mitigate heterogeneity by assigning varying LoRA depths based on devices' capabilities. This strategy, however, overlooks the potential importance of weaker devices in FedFT \cite{chen2024heterogeneity}: assigning them smaller LoRA depths can significantly degrade fine-tuning performance. Moreover, these FedFT approaches are typically initialized with off-the-shelf pre-trained models without synthetic data augmentation, indicating that convergence performance could be further improved \cite{karimireddy2021breaking}.

Recent theoretical work \cite{jhunjhunwala2025initialization} has shown that a carefully initialized pre-trained global model can mitigate the adverse effects of non-IID data and accelerate convergence in FedFT. However, models pre-trained on public datasets (\eg, ImageNet \cite{deng2009imagenet}) may not reflect the device-specific data distributions \cite{chen2023on}, making them suboptimal as initialization points for FedFT. Several studies \cite{zhang2025gpt,zhang2023federated} have introduced diffusion models (DM) \cite{sohl2015deep} to generate synthetic images that approximate the label distribution on each device. These synthetic images are then used to pre-train a global model, serving as the initialization for FedFT. However, these approaches rely solely on text-conditioned synthesis (\eg, labels or text embeddings) and may fail to capture the unique feature distribution of each device \cite{zhou2024federated}, leading to a style gap between synthetic and real data that diminishes the effectiveness of augmentation.

Overall, none of these approaches adequately addresses the challenges outlined above, summarized in Table \ref{tbl:FLcomparison}. More importantly, these approaches treat synthetic pre-training and FedFT as \textit{isolated} stages, and ignore their synergy: insufficient images weaken pre-training effectiveness for FedFT, whereas excessive images introduce substantial computation overhead, thereby prolonging generation and pre-training time. Therefore, \textit{the key challenge lies in determining the optimal number of generated images, so as to balance pre-training benefit for FedFT against generation/pre-training cost}.

\textbf{Overview of the Proposed Approach.} To this end, we propose FeDiSyn, a unified framework that holistically considers pre-training and FedFT to minimize the overall completion time for LVM training, which is built upon ``one law, two optimizations''. \textit{First}, we introduce a scaling law for synthetic pre-training under FedFT setting, which captures the relationship among accuracy, total training time, and the synthetic data scaling factor, providing quantitative guidance for determining the optimal number of generated images. \textit{Second}, for pre-training optimization, each device extracts features by encoding local images into latents, which are then uploaded to the server. By combining these latents with semantically enhanced labels via a DM, the server generates synthetic images to pre-train a global model. Since the synthetic images capture cross-device label and feature distributions, the pre-trained model yields a superior initialization for FedFT, effectively addressing both label and feature distribution skews. \textit{Third}, for FedFT optimization, FeDiSyn introduces a contribution-aware LoRA configuration and bandwidth allocation algorithm. By configuring potentially important devices sufficient LoRA depth and balancing completion times via bandwidth allocation, the FedFT performance can be further improved while addressing system heterogeneity.

Our main contributions are summarized as follows.
\begin{itemize}
    \item We propose FeDiSyn, a unified framework that holistically considers the close interplay between pre-training and FedFT to reduce the overall LVM training time. To our knowledge, this is the first work to explore the scaling law for synthetic pre-training under FedFT setting, and systematically optimize these two stages in a unified manner.
    \item We develop a synthetic pre-training method that tackles non-IID data from both label and feature distribution perspectives. In addition, we design a contribution-aware LoRA configuration and bandwidth allocation algorithm to improve FedFT performance while addressing system heterogeneity.
    \item We implement the proposed FeDiSyn framework and related algorithms on a real-world testbed. Experimental results on various models and datasets show that the proposed solution outperforms the state-of-the-art solutions.
\end{itemize}

\section{Background and Motivation}\label{sec:related}

\subsection{Pre-Training Model Initialization for FedFT}

\textbf{The Importance of Pre-Training and Initialization.} As a central challenge in FedFT, non-IID data is typically addressed through various aspects, such as device sampling \cite{chen2024heterogeneity,ma2024feduc} and aggregation mechanism \cite{ma2021fedsa,ma2025air}, while the impact of model initialization is often overlooked in most studies. Several recent works have highlighted the critical importance of pre-training for the initial model in FedFT. Nguyen \etal \cite{nguyen2022where} observe that models initialized with pre-trained weights typically exhibit significantly lower training loss values at initialization compared to those initialized randomly. Karimireddy \etal \cite{karimireddy2021breaking} further demonstrate that this lower initial loss resulting from pre-training facilitates faster convergence during subsequent FedFT. Chen \etal \cite{chen2023on} note that pre-training shapes a better conditioned loss surface and steers the model into a location closer to the optimal solution, enhancing the stability of model aggregation. These phenomena are theoretically supported by \cite{jhunjhunwala2025initialization}, which proves that pre-training effectively reduces sensitivity to certain hyperparameters within FedFT by diminishing misaligned filters at initialization.

\textbf{Synthetic Image for Pre-Training LVMs.} Although pre-training initialization can enhance FedFT performance, models pre-trained on public datasets (\eg, ImageNet) show limited gains due to domain mismatch between the public data and the device-specific data. For example, in chest X-ray analysis, an ImageNet pre-trained model achieves only 0.2\% accuracy improvement over random initialization, whereas a domain-relevant synthetic image pre-trained model yields a 5.9\% gain \cite{chen2023on}. To address this, GPT-FL \cite{zhang2025gpt} and FGL \cite{zhang2023federated} upload labels or text descriptions of device datasets to the server for synthetic image generation. However, these approaches rely solely on text-conditioned generation, which fails to capture device-specific feature distributions. As a result, the synthetic images may not align well with real local data, limiting their improvement for FedFT.

\begin{figure}[t]
\centering
\includegraphics[width=0.80\linewidth]{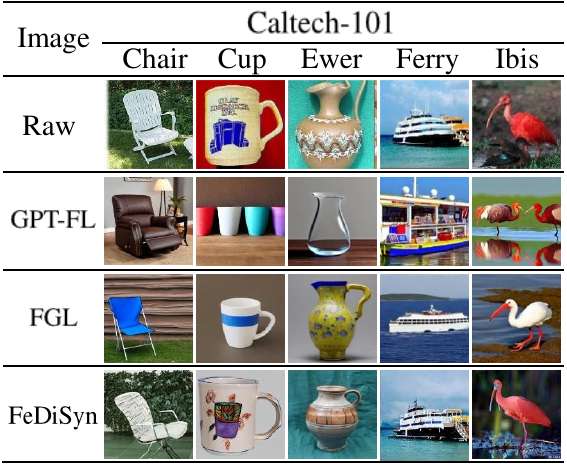}
\vspace{-3mm}
\caption{Comparison of FeDiSyn and other synthetic image generation methods on Caltech-101.}
\label{fig:example}
\vspace{-5mm}
\end{figure}

In contrast, FeDiSyn uploads latents encoding local data features to generate synthetic images, as elaborated in Section \ref{sec:framework}. For example, generated images for Caltech-101 \cite{li2004Learning} in Fig. \ref{fig:example} show that FeDiSyn can obtain synthetic images that better align with the feature distributions of raw data compared to GPT-FL and FGL.

\subsection{Scaling Law of Synthetic-Image Pre-Training for FedFT}

\begin{figure}[t]
\begin{minipage}[t]{0.48\linewidth}
\includegraphics[width=\linewidth]{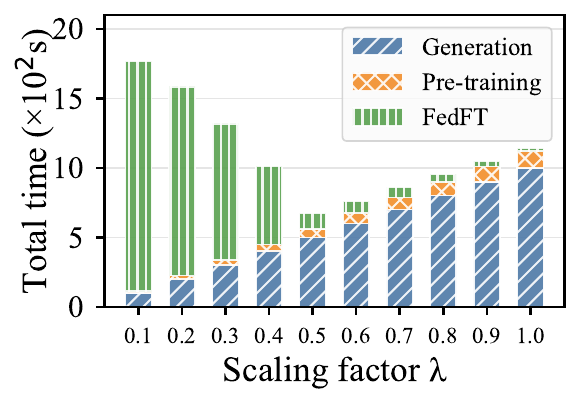}
\vspace{-3mm}
  \small{\caption{The effect of different $\lambda$ on total time to reach 70\% accuracy on Caltech-101.} \label{fig:diff_lambda_caltech101_total_time}}
\end{minipage}%
\hspace{2mm}
\begin{minipage}[t]{0.48\linewidth}
\includegraphics[width=\linewidth]{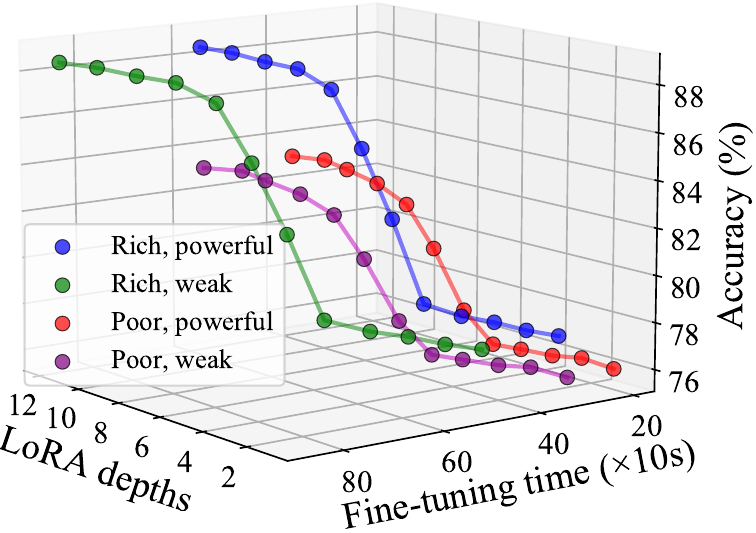}
\vspace{-3mm}
  \small{\caption{The tradeoff between heterogeneity and performance.} \label{fig:lora_tradeoff}}
\end{minipage}%
\vspace{-5mm}
\end{figure}

Following the groundbreaking success of large-scale models, scaling laws have emerged as a critical framework for quantifying the power-law relationships between model/data scale and performance \cite{kaplan2020scaling}. Prior studies have demonstrated that pre-training data volume has a significant impact on model performance \cite{zhang2024when,pearce2025scaling}, which has motivated recent extensions of these principles to synthetic data, exploring how generative data scale influences convergence \cite{fan2024scaling,kang2025demystifying}. However, these investigations focus exclusively on centralized training paradigms and are therefore not directly applicable to the FedFT setting, where the capabilities of servers for pre-training and edge devices for fine-tuning jointly determine the overall training time. As illustrated in Fig. \ref{fig:diff_lambda_caltech101_total_time} on Caltech-101, the synthetic scaling factor $\lambda$ introduces a trade-off in FedFT: increasing $\lambda$ accelerates convergence but incurs substantial overhead in image generation and pre-training. Therefore, the total completion time follows a U-shaped curve, which, in this example, attains its minimum near $\lambda = 0.5$.

To provide guidance for synthetic pre-training in the FedFT setting, we propose a novel scaling law that quantifies the relationship between synthetic volume, target accuracy, and total training time, enabling the analytical determination of the optimal scaling factor $\lambda$, as elaborated in Section \ref{sec:algorithm}.

\subsection{Federated Fine-Tuning with LoRA for LVMs}\label{subsec:importance_lora_depth}

LoRA \cite{hu2022lora} fine-tuning is a specific PEFT approach via low-rank matrix decomposition. In LoRA-based FedFT, each device only fine-tunes and uploads the lightweight inserted LoRA layers while keeping the pre-trained model frozen. The server aggregates the uploaded LoRA parameters and then distributes them back to all devices. Most existing LoRA-based FedFT approaches for LVMs \cite{wang2025federated,su2024fedra} insert LoRA layers into all transformer layers, causing significant computation overhead. To reduce this burden, a few studies \cite{pfeiffer2024efficient,liu2025adaptive} insert LoRA only to a subset of layers, enabling more efficient fine-tuning on resource-constrained devices. Futhermore, these works typically address system heterogeneity by assigning LoRA depths based solely on device capabilities: powerful devices receive deeper LoRA layers, while resource-constrained devices get smaller ones. However, weak devices may play a critical role in FedFT, and assigning them shallow LoRA depths can significantly degrade performance. We conduct a case study with ViT-B/16 fine-tuned on CIFAR-10 \cite{krizhevsky2009learning} in Fig. \ref{fig:lora_tradeoff}, which compares accuracy and fine-tuning time across four device types: (i) rich-data/powerful-capability, (ii) rich-data/weak-capability, (iii) poor-data/powerful-capability, and (iv) poor-data/weak-capability.
As shown, powerful but data-scarce devices (red line) gain little benefit from deeper LoRA depths, whereas data-rich devices with weak-capability (green line) achieve accuracy comparable to powerful ones (blue line) when given a deeper LoRA, albeit with longer fine-tuning time. These observations highlight a trade-off: It is worth setting a larger LoRA depth for resource-constrained yet critical devices, but these devices may become stragglers that delay the overall FedFT process.

To this end, we propose a contribution-aware LoRA configuration and bandwidth allocation strategy in FeDiSyn. LoRA depth for each device is determined by its expected performance gain, while bandwidth is allocated to balance device completion times, effectively addressing system heterogeneity. Details are elaborated in Section \ref{sec:algorithm}.

\section{System Design of FeDiSyn}\label{sec:framework}

\subsection{Preliminaries}\label{subsec:reliminaries}

We consider a $K$-class classification problem performed through FedFT with label space $\mathcal{C} = \{c_1, c_2, ..., c_K\}$. The system contains a centralized server with relatively abundant computational resources and a set of devices $\mathcal{V} = \{v_1, v_2, ..., v_{N}\}$. Each device $v_i$ owns a local dataset $\mathcal{D}_i$, with size $D_i=|\mathcal{D}_i|$. The global dataset across all devices is represented as $\mathcal{D}$, with size $D=|\mathcal{D}|= \sum_{v_i\in \mathcal{V}}D_i$. The data labeled as $c_k$ on $v_i$ is denoted as $\mathcal{D}_i^k$ with size $D_i^k=|\mathcal{D}_i^k|$, and it holds that $\sum_{c_k\in\mathcal{C}}D_i^k = D_i$. The loss function of $v_i$ is defined as
\begin{equation}\label{eq:filossfuction}
f_i(\mathbf{w})\triangleq\tfrac{1}{D_i}\sum\nolimits_{\xi\in \mathcal{D}_i}f_i(\mathbf{w};\xi)\mbox{,}
\end{equation}
where $\mathbf{w}$ is the parameter vector, and $f_i(\mathbf{w};\xi)$ is the loss over sample $\xi$ in $\mathcal{D}_i$. The global loss function on all the distributed datasets is
\begin{equation}\label{eq:lossfuction}
F(\mathbf{w})\triangleq\sum\nolimits_{v_i\in \mathcal{V}}\tfrac{D_i}{D}f_i(\mathbf{w})\mbox{.}
\end{equation}
The learning problem is to find the optimal parameter vector $\mathbf{w}^*$ so as to minimize $F(\mathbf{w})$, \ie, $\mathbf{w}^*=\mathop{\arg\min}_{\mathbf{w}} F(\mathbf{w})$.

\subsection{System Overview}\label{subsec:overview}

\begin{figure}[t]
\centering
\includegraphics[width=1\linewidth]{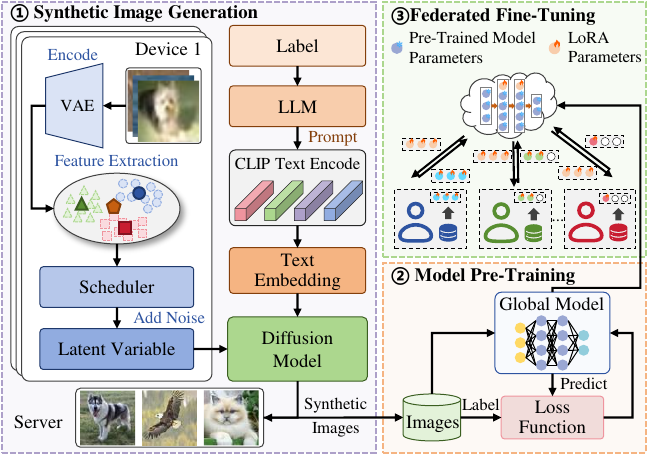}
\vspace{-6mm}
\caption{Overview of FeDiSyn.}
\label{fig:overview}
\vspace{-5mm}
\end{figure}

As shown in Fig. \ref{fig:overview}, the process of FeDiSyn consists of three steps. \textbf{\large{\textcircled{\small{1}}}} \textbf{Synthetic image generation:} Devices encode private images into low-dimensional latents using a lightweight variational autoencoder (VAE). These latents are clustered to obtain class-specific centroids and a global average representation, which are then sent to the server with their labels. The server uses an large language model (LLM) to expand labels into diverse text prompts, encodes them via CLIP's \cite{radford2021learning} text encoder, and feeds the resulting embeddings along with the uploaded latents into a DM (\eg, Stable Diffusion) to generate synthetic images, where latents govern stylistic features (\eg, lighting, texture) and text embeddings condition semantics (\eg, class-specific details).
\textbf{\large{\textcircled{\small{2}}}} \textbf{Model pre-training:} The server pre-trains a global model over synthetic images, yielding an initial model that captures cross-device features for subsequent FedFT.
\textbf{\large{\textcircled{\small{3}}}} \textbf{Federated fine-tuning:} The initialized model is sent to devices for LoRA-based fine-tuning under the FedFT framework. Specifically, each device keeps the pre-trained model frozen and updates only LoRA parameters, which are aggregated by the server and redistributed until convergence.
The following sections detail each step.

\subsection{Synthetic Image Generation}\label{subsec:generation}

\textbf{Data Feature Extraction.} To align synthetic images with device-specific styles while preserving privacy, each device $v_i$ extracts the features of its raw images locally in the latent space. Specifically, each sample $\mathcal{D}_{i,j}^k\in \mathcal{D}_i^k$ of class $c_k$ on device $v_i$ is encoded by a VAE $\mathcal{E}$ to obtain latents $e_{i,j}^k=\mathcal{E}(\mathcal{D}_{i,j}^k)$.
Subsequently, for the same class $c_k$, device $v_i$ clusters the latents $e_{i,j}^k, \forall j$ into $Z$ cluster centroids $\mathbf{e}_{i,z}^k,z\in [1,Z]$, and calculate their average representation as $\mathbf{e}_{i}^k$. $\mathbf{e}_{i,z}^k$ and $\mathbf{e}_{i}^k$ capture the style characteristics of the device distribution \cite{yang2024exploring}.
Then the device uses a scheduler to add noise to the encoded latents at a randomly chosen diffusion timestep $s\in\left\{0,...,S\right\}$ \cite{rombach2022high} as
\begin{equation}
\label{eqn:addnosie}
     \begin{aligned}
    &\bar{\mathbf{e}}_{i,z}^{k,s}=\sqrt{\alpha_s}\mathbf{e}_{i,z}^k+\sqrt{1-\alpha_s}\epsilon, \epsilon\sim \mathcal{N}(0, I), z\in[1,Z], \\
    &\bar{\mathbf{e}}_{i}^{k,s}=\sqrt{\alpha_s}\mathbf{e}_{i}^k+\sqrt{1-\alpha_s}\epsilon, \epsilon\sim \mathcal{N}(0, I),
    \end{aligned}
\end{equation}
where $\alpha_s$ controls noise intensity at timestep $s$. The device retains the raw data locally and only uploads the cluster centroids $\bar{\mathbf{e}}_{i,z}^{k,S}$, the average representation $\bar{\mathbf{e}}_{i}^{k,S}$, and its class label $c_k$ to the server.

\textbf{Label Collection and Prompt Engineering.} The server collects the data volume $D_i^k$ labeled as $c_k$ of each device $v_i$ and calculates the total amount of data labeled as $c_k$ across all devices by $D^k = \sum_{v_i \in \mathcal{V}} D_i^k$. To enrich semantic variety, a deployed LLM expands each label into multiple descriptive prompts. It has been verified that language enhancement is an effective way for enlarging synthetic image diversity for pre-training \cite{he2022synthetic}. For example, for the label ``dog'', the prompt can be semantically expanded as follows:

Q: Write a sentence using the word ``dog'' to describe the characteristics and actions of the object.

A: A golden retriever playing on the grass.

These prompts are encoded into text embeddings via CLIP's text encoder, ensuring semantic alignment between text and image latent spaces. To preserve the real data distribution, the synthetic image size for class $c_k$ is set as $\hat{D}^k = \lambda D^k$. $\lambda$ is the scaling factor, which is determined according to our proposed scaling law in Section \ref{sec:algorithm}.

\textbf{Synthetic Image Generated by Diffusion Model.} On the server, for each class $c_k$, the received cluster centroids $\bar{\mathbf{e}}_{i,z}^{k,S}$ and average representation $\bar{\mathbf{e}}_{i}^{k,S}$ are weighted as $\bar{E}_{i,z}^{k,S}= \rho \bar{\mathbf{e}}_{i,z}^{k,S} + (1-\rho)\bar{\mathbf{e}}_{i}^{k,S}$, where $\rho$ is a factor with a value between 0 and 1. Next, each latent $\bar{E}_{i,z}^{k,S}$ is paired with a text embedding and fed into a latent DM, where latents guide stylistic consistency and text embeddings enforce data diversity. At each reverse step, a PNDM scheduler \cite{liu2022pseudo} is employed for sampling, iteratively denoising $\bar{E}_{i,z}^{k,S}$ over $S$ steps via
\begin{equation}
\label{eqn:denosie}
    \phi(\bar{E}_{i,z}^{k,S},\epsilon_s,s,s-\delta)=\tfrac{\sqrt{\overline{\alpha}_{s-\delta}}}{\sqrt{\overline{\alpha}_s}}\bar{E}_{i,z}^{k,S}
    -\tfrac{(\overline{\alpha}_{s-\delta}-\overline{\alpha}_s)}{\sqrt{\overline{\alpha}_s}\left(\sqrt{(1-\overline{\alpha}_{s-\delta})\overline{\alpha}_s}+\sqrt{(1-\overline{\alpha}_s)\overline{\alpha}_{s-\delta}}\right)}\epsilon_s,
\end{equation}
where $\delta$ is the step interval, and the gradient term $\epsilon_s$ is computed using numerical methods such as the Runge-Kutta method or the linear multi-step method. Ultimately, all $\bar{E}_{i,z}^{k,S}, \forall i,k$ undergo a certain number of denoising process to produce synthetic images $\hat{\mathcal{D}}$ with size $\hat{D}=\sum_{c_k\in\mathcal{C}}\hat{D}_k=\lambda\sum_{c_k\in\mathcal{C}} D_k$ at server side.

\begin{remark}[Privacy Analysis]
Transmitting latent representations is a widely used strategy for data augmentation in FL. Considering the limited amount of latent information uploaded by each device, our approach introduces minimal privacy leakage. Furthermore, similar to \cite{yang2024exploring}, we generate images using DM from noise-added features. As a result, even when the original images contain sensitive content (\eg, text or faces), the generated images exhibit entirely different semantic details. Therefore, despite potential visual similarities, they do not reveal or compromise the underlying private information.
\end{remark}

\subsection{Model Pre-Training over Synthetic Image}\label{subsec:pretraining}

With the generated synthetic images $\hat{\mathcal{D}}$ on the server, a global model is pre-trained in a standard supervised fashion. Similar to Eq. \eqref{eq:lossfuction}, we denote $\hat{F}(\mathbf{w})$ as the loss function over the synthetic images $\hat{\mathcal{D}}$ for model parameters $\mathbf{w}$. Starting from an initial parameter vector $\hat{\mathbf{w}}^0$, the server performs $\hat{T}$ iterations of model pre-training by
\begin{equation}
\label{eqn:synloss}
    \hat{\mathbf{w}}^t=\hat{\mathbf{w}}^{t-1}-\eta\nabla\hat{F}(\hat{\mathbf{w}}^{t-1}),
\end{equation}
where $\eta$ is the learning rate, and $\nabla$ is the gradient operator. The global model after $\hat{T}$ iterations of pre-training is denoted by $\overline{\mathbf{w}}=\hat{\mathbf{w}}^{\hat{T}}$, which acts as the initial model for the subsequent FedFT process.

\subsection{Federated Fine-Tuning with LoRA}\label{subsec:fftlora}

\begin{figure}[t]
\centering
\includegraphics[width=\linewidth]{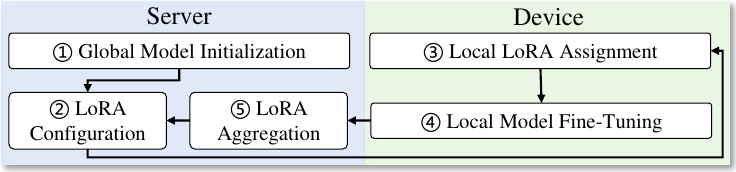}
\vspace{-7mm}
\caption{Federated fine-tuning with LoRA.}
\label{fig:lora_alg}
\vspace{-4mm}
\end{figure}

After the pre-training model $\overline{\mathbf{w}}$ is obtained, FeDiSyn performs FedFT with LoRA for the LVM through the five phases shown in Fig. \ref{fig:lora_alg}.

\textbf{\large{\textcircled{\small{1}}}} \textbf{Global Model Initialization.} The server first injects the LoRA layers into the pre-trained model $\overline{\mathbf{w}}$, \ie, embedding initial LoRA matrices $\widetilde{\mathbf{w}}^{0}=\{\widetilde{\mathbf{w}}_l^{0}, \forall l\in [0,L-1]\}$ within $L$ transformer layers of the neural network. Then the embedded pre-trained model $\mathbf{w}^0=\{\overline{\mathbf{w}}, \widetilde{\mathbf{w}}^0\}$ is distributed to all devices in $\mathcal{V}$.

\textbf{\large{\textcircled{\small{2}}}} \textbf{LoRA Configuration.} Prior studies \cite{zhang2023adaptive,liu2025adaptive} show that inserting LoRA into deeper layers yields better fine-tuning performance than shallow layers. Therefore, we define LoRA depth as the number of layers with LoRA modules, counted from the deepest output layer upward. At each round $t$, the server assigns each device $v_i$ a depth $x_i^t$ based on its heterogeneous resources, thus only the last $x_i^t$ layers $L_i^t=\{l|L-x_i^t \le l < L\}$ are fine-tuned. The LoRA configuration strategy for all devices at round $t$ is denoted as $\mathcal{X}^t = \{x_i^t|v_i\in\mathcal{V}\}$, which will be described in Section \ref{sec:algorithm}.

\textbf{\large{\textcircled{\small{3}}}} \textbf{Local LoRA Assignment.} At each round $t$, the server distributes the configuration information $\mathcal{X}^t$ and the global LoRA parameters $\widetilde{\mathbf{w}}^t=\{\widetilde{\mathbf{w}}_{l}^{t}| \forall l\in [0,L-1]\}$ to all devices in $\mathcal{V}$. The $l$-th LoRA layer of device $v_i$'s local model is assigned as
\begin{equation}
\label{eqn:update}
    \widetilde{\mathbf{w}}_{i,l}^{t}=\widetilde{\mathbf{w}}_l^{t}.
\end{equation}

\textbf{\large{\textcircled{\small{4}}}} \textbf{Local Model Fine-Tuning.}
After the local LoRA configuration, $v_i$ fine-tunes the local model $\mathbf{w}_i^t=\{\overline{\mathbf{w}}, \widetilde{\mathbf{w}}_i^t\}$ on its local dataset $\mathcal{D}_i$. The update of each LoRA layer $\widetilde{\mathbf{w}}_{i,l}^{t}$ at round $t$ is expressed as
\begin{equation}
\label{eqn:lora-fine-tuning}
    \begin{aligned}
     \widetilde{\mathbf{w}}_{i,l}^{t+1}=
    \begin{cases}
    \widetilde{\mathbf{w}}_{i,l}^{t}-\eta\nabla f_i(\widetilde{\mathbf{w}}_{i,l}^{t}), & L - x_i^t \le l < L \\
    \widetilde{\mathbf{w}}_{i,l}^{t}, & \mbox{otherwise}
    \end{cases}.
    \end{aligned}
\end{equation}
where $\eta$ is the learning rate and $\nabla$ is the gradient operator.

\textbf{\large{\textcircled{\small{5}}}} \textbf{LoRA Aggregation.} After local fine-tuning, each device $v_i$ uploads the fine-tuned LoRA parameters $\widetilde{\mathbf{w}}_{i,l}^{t+1}, L - x_i^t \le l < L$ to the server. Upon receiving the updated LoRA parameters from all devices, for each layer $l$, the server aggregates them as
\begin{equation}
\label{eqn:aggregation}
    \widetilde{\mathbf{w}}_l^{t+1}=\frac{\sum_{v_i\in\mathcal{V}}[l\ge L-x_i^t] D_i\widetilde{\mathbf{w}}_{i,l}^{t+1}}{\sum_{v_i\in\mathcal{V}}[l\ge L-x_i^t] D_i},
\end{equation}
where $[\cdot]$ is the Iverson notation. As a result, the global LoRA parameters are updated as $\widetilde{\mathbf{w}}^{t+1}=\{\widetilde{\mathbf{w}}_{l}^{t+1}| \forall l\in [0,L-1]\}$, and the server configures LoRA for the next round $t+1$. The process repeats until the model converges or reaches satisfactory accuracy.

\section{Convergence Analysis}\label{sec:convergence}

We make the following assumptions on the loss functions $F(\mathbf{w})$ for convergence analysis, which are widely used in prior works \cite{wang2019adaptive}.
\begin{assumption}[Lipschitz]\label{ass:smoothness} The loss function $F$ has a $\mathcal{L}$-Lispschitzian gradient for $\forall \mathbf{w}_1,\mathbf{w}_2$, \ie, $ \|\nabla F(\mathbf{w}_1)-\nabla F(\mathbf{w}_2)\| \le \mathcal{L} \|\mathbf{w}_1-\mathbf{w}_2\|$.
\end{assumption}

\begin{assumption}[Gradient Bounded]\label{ass:bounded}
The expected squared norm of gradients is uniformly bounded for $\forall\mathbf{w}_i^t$, \ie, $\mathbb{E}\|\nabla F(\mathbf{w}_i^t)\|^2\le G^2$.
\end{assumption}
For LoRA fine-tuning, each LoRA matrix is further decomposed as $\widetilde{\mathbf{w}}_i^{t}=\mathcal{B}_i^t\mathcal{A}_i^t, \forall i,t$, where $\mathcal{B}_i^t$ and $\mathcal{A}_i^t$ are low-rank matrices.
\begin{assumption}[LoRA Bounded]\label{ass:lorabounded}
There exist constants $C_\mathcal{A},C_\mathcal{B}>0$, such that $\|\mathcal{A}_i^t\|_2\le C_\mathcal{A}$, $\|\mathcal{B}_i^t\|_2\le C_\mathcal{B}$.
\end{assumption}

Based on the above assumptions, we derive the convergence bound of FeDiSyn by the following theorem.
\begin{Theorem}[Convergence of FeDiSyn]
\label{the:convergence}
After $T$ rounds of fine-tuning in FeDiSyn, the expectation of the gradient of the LoRA matrices satisfies
\begin{align}
    &\frac{1}{TN}\sum\nolimits_{t=1}^T\sum\nolimits_{v_i\in\mathcal{V}}\left(\mathbb{E}\|\nabla F(\mathcal{A}_i^t)\|^2+\mathbb{E}\|\nabla F(\mathcal{B}_i^t)\|^2\right)\notag\\&\le\frac{1}{N}\sum\nolimits_{v_i\in\mathcal{V}}\frac{\mathbb{E}[F(\mathbf{w}^0)]-\mathbb{E}[F(\mathbf{w}_i^{*})]}{\eta T}+\kappa,
\end{align}
where $\mathbf{w}_i^*$ denotes the optimal model on $v_i$, $\kappa$ is a constant satisfying $[G^3C_\mathcal{A}C_\mathcal{B}+\frac{3\mathcal{L}}{2}(G^2C_\mathcal{A}^2+G^2C_\mathcal{B}^2+\eta^2G^4C_\mathcal{A}^2C_\mathcal{B}^2)]\eta+\frac{G^2}{2}\le\kappa$.
\end{Theorem}
Due to limited space, we omit the proof of Theorem \ref{the:convergence} here.
\begin{remark}[LoRA-based FedFT Convergence]
It is noticed that, unlike prior convergence analyses for FedFT \cite{liu2025adaptive,wang2025federated}, which typically study the convergence behavior only at the level of the overall model parameter $\mathbf{w}$, our analysis is developed specifically for LoRA-based FedFT and explicitly characterizes the convergence of the two low-rank matrices $\mathcal{A}$ and $\mathcal{B}$. As a result, our convergence bound provides a fine-grained theoretical understanding of the training in LoRA-based FedFT, and better reveals how the low-rank decomposition influences convergence.
\end{remark}

\section{Problem Formulation and Algorithm Description}\label{sec:algorithm}

\subsection{Problem Formulation}\label{subsec:formulation}

The total completion time of FeDiSyn includes the image generation, pre-training, and FedFT time. Let $\hat{h}$ denote the time to generate one image on server, then the total image generation time is expressed as $H^{\rm gene}=\hat{D}\hat{h} =\lambda D\hat{h}$.
Next, the server pre-trains LVM on synthetic data for $\hat{T}$ rounds, and the pre-training time can be expressed as $H^{\rm pre}=\lambda D \tilde{h}\hat{T}$, where $\tilde{h}$ is the single round training time for one image on server.

For FedFT, although deeper LoRA layers improve its performance, practical deployments are constrained by resource limitations such as computation, communication, and memory.
Let $\overline{h}_i$ denote the time for data loading and performing a full forward pass through the LVM on $v_i$ per sample, and $h_i$ denote the additional time for backpropagating through per LoRA layer. Since the computation capacity of $v_i$ is fixed, both $\overline{h}_i$ and $h_i$ can be measured in advance. Therefore, the fine-tuning time on device $v_i$ at round $t$ is given by $H_i^{t, \rm comp}= D_i(\overline{h}_i + x_i^t h_i)$. The transmission time primarily arises from uploading LoRA parameters via FDMA, which allows simultaneous transmissions from multiple devices \cite{wang2025federated}. Let $B_i^t$ denote the bandwidth allocated to $v_i$ at round $t$, then the uplink transmission rate $r_i^t=B_i^t\log(1+\frac{p_ih_i^t}{\sigma^2})$ is modeled by the Shannon capacity, where $p_i$ is the transmission power, $h_i^t$ is the channel gain, and $\sigma^2$ is the noise power. Therefore, the transmission time for $v_i$ at round $t$ is $H_i^{t,\rm comm}=\tfrac{\tau x_i^t}{r_i^t}$, where $\tau$ is the parameter size per LoRA layer. The duration of round $t$ is determined by the slowest device, \ie,
\begin{equation}
\label{eqn:max_time}
H^t=\max\nolimits_{v_i \in\mathcal{V}}\{H_i^{t,\rm comp}+H_i^{t,\rm comm}\}.
\end{equation}

For memory, let $\overline{m}$ denote the fixed memory to load the LVM and store forward activations, and $m$ represent the extra memory incurred by gradient storage per LoRA layer during backpropagation. Then, the total memory overhead for fine-tuning on $v_i$ at round $t$ is
\begin{equation}
\label{eqn:memory}
    m_i^t = \overline{m} + x_i^t m.
\end{equation}

Let the total completion time $H=H^{\rm gene}+H^{\rm pre}+\sum\nolimits_{t=1}^T H^t$, we formulate the problem as follows:
\begin{subequations}
\begin{align}
\textbf{(P1)}:&\min\nolimits_{\lambda,\mathcal{X}^t,\mathcal{B}^t} H \\
{\st} \quad &F(\mathbf{w}^T)-F(\mathbf{w}^*)\le \epsilon \label{eqn:convergencethreshold}\\
& \sum\nolimits_{v_i\in\mathcal{V}} B_i^t \le B &\forall t\in[0,T] \label{eqn:bandwidthcons}\\
& m_i^t \le M_i & \forall i\in [1,N] \label{eqn:memorycons}
\end{align}
\end{subequations}
Constraint \eqref{eqn:convergencethreshold} represents the global model converges after $T$ rounds, where $\epsilon$ is the convergence threshold.
Constraint \eqref{eqn:bandwidthcons} ensures the total allocated bandwidth does not exceed the budget $B$ in each round. Constraint \eqref{eqn:memorycons} ensures each device $ v_i$'s memory usage does not exceed its available memory $M_i$.
Our objective is to decide the data scaling factor $\lambda$, the LoRA depth configuration $\mathcal{X}^t=\{x_i^t|v_i\in\mathcal{V}\}$ and bandwidth allocation $\mathcal{B}^t=\{B_i^t|v_i\in\mathcal{V}\}$ for each round, minimizing the total completion time, \ie, $\min_{\lambda,\mathcal{X}^t,\mathcal{B}^t}H$.

\begin{algorithm}[tb]
\caption{Diffusion Model-Driven Synthetic Image Pre-Training for Federated Fine-Tuning (FeDiSyn)}
\label{alg:FeDiSyn}
\begin{algorithmic}[1]
\REQUIRE Data volume $D_i$, available memory $M_i$, bandwidth budget $B$\\
\ENSURE The fine-tuned model $\mathbf{w}^T$\\
\STATE/* \textit{Step 1. Determining Scaling Factor.}*/
\STATE Conduct a set of small-scale experiments to measure $H(\lambda, Acc)$
\STATE Fit coefficients $\kappa_1$-$\kappa_4$ using nonlinear least squares method
\STATE Obtain optimal scaling factor $\lambda^*=\sqrt[\kappa_2 + 1]{\tfrac{\kappa_2e^{\kappa_3 Acc}}{h_0\kappa_1}}$\label{alg1:get lambda}
\STATE/* \textit{Step 2. Generating Synthetic Images and Pre-training.}*/
\STATE Obtain the centroids $\mathbf{e}_{i,z}^k$ and average representation $\mathbf{e}_i^k$ on $v_i$\label{alg1:cluster}
\STATE Devices upload noisy latents $\bar{\mathbf{e}}_{i,z}^{k,s}$, $\bar{\mathbf{e}}_{i}^{k,s}$ and labels of each $c_k$\label{alg1:upload latents}
\STATE Server expands the labels $\mathcal{C}$ into prompts\label{alg1:prompts}
\STATE Server performs image generation with $\lambda^*$ yields $\hat{\mathcal{D}}$\label{alg1:generation}\\
\STATE Pre-train LVM over $\hat{\mathcal{D}}$ and distribute LoRA-inserted models\label{alg1:pre-train}\\
\STATE/* \textit{Step 3. Federated Fine-tuning.}*/
\FOR{each $t\in T$}\label{alg1:begin t}
\STATE/* \textit{Contribution-aware LoRA configuration and bandwidth allocation algorithm (CLCBA).}*/
    \FOR{each $v_i \in \mathcal{V}$}\label{alg1:begin initial}
        \STATE $x_i^t \gets 1$ , $x_i^{\rm max} \gets \min\left(\left\lfloor (M_i-\overline{m}) / m\} \right\rfloor, L\right)$    \label{alg1:set x_max}
    \ENDFOR
    \STATE $R_{\rm max}^t \gets 0$ \label{alg1:end initial}
    \WHILE{\TRUE} \label{alg1:begin while}
        \FOR{each $v_i \in \mathcal{V}$}
            \IF{$x_i^t < x_i^{\rm max}$}
                \STATE $x_i^t \gets x_i^t+1$
                \STATE Obtain $\mathcal{B}^t$ under $\mathcal{X}^t$ by water-filling based bandwidth allocation sub-algorithm (WFBA, Alg. \ref{alg:WFBA})
                \STATE Calculate completion time $H^t$ by Eq. \eqref{eqn:max_time}
                \STATE Calculate performance gain $R^t$ by Eq. \eqref{eqn:reward}
                \STATE $\Delta R_i^t \gets R^t-R_{\rm max}^t$, $x_i^t \gets x_i^t-1$
            \ENDIF
        \ENDFOR
        \IF{$\exists \Delta R_i^t > 0$}
            \STATE $i \gets { \rm argmax}_{v_i \in \mathcal{V}} \Delta R_i^t$, $R_{\rm max}^t \gets R_{\rm max}^t+\Delta R_i^t$, $x_i^t \gets x_i^t+1$
        \ELSE
            \STATE \textbf{break}
        \ENDIF
    \ENDWHILE   \label{alg1:end while}
    \STATE Obtain $\mathcal{B}^t$ under $\mathcal{X}^t$ by Alg. \ref{alg:WFBA}
    \STATE After LoRA assignment, devices perform local fine-tuning
    \STATE Server aggregates the LoRA matrices uploaded by devices
\ENDFOR \label{alg1:end t}
\RETURN The fine-tuned model $\mathbf{w}^T$
\end{algorithmic}
\end{algorithm}

\subsection{Algorithm Description}\label{subsec:algorithm description}

In this section, we introduce the algorithm description for our FeDiSyn to solve problem \textbf{P1}, as described in Alg. \ref{alg:FeDiSyn}.

\textbf{\large{\textcircled{\small{1}}}}  \textbf{Determining Scaling Factor.} The key challenge of solving problem \textbf{P1} is determining the scaling factor $\lambda$ to balance the time required for synthetic image generation/pre-training and FedFT.
From Section \ref{subsec:formulation}, the total time for generation and pre-training is $H^{\rm gene}+H^{\rm pre}=\lambda D\hat{h}+\lambda D \tilde{h}\hat{T}=\lambda(\hat{h}+\tilde{h}\hat{T})D=\lambda h_0$, where $h_0\triangleq(\hat{h}+\tilde{h}\hat{T})D$ is defined as the unit time for generation and pre-training. Conversely, empirical analysis across multiple datasets suggests that the subsequent FedFT time is negatively correlated with $\lambda$ and grows exponentially with the target accuracy $Acc$. To quantify this relationship, we propose the following \textit{scaling law} to describe the total completion time as
\begin{equation}
\label{eqn:time_lambda_acc}
     H(\lambda, Acc)=h_0\lambda +\tfrac{1}{\kappa_1\lambda^{\kappa_2}}e^{\kappa_3 Acc}+\kappa_4,
\end{equation}
where $h_0 \lambda$ represents the linear growth of generation and pre-training time, and the remaining terms capture the diminishing returns of synthetic data on FedFT acceleration. The  architecture-specific coefficients $\kappa_1$-$\kappa_4$, which depend on the dataset, model and device capacity, can be estimated in a short calibration phase using a set of small-scale pilot runs, detailed in Section \ref{subsec:result}. To find the optimal scaling factor $\lambda^*$ that minimizes the total training time, we take the derivative of Eq. \eqref{eqn:time_lambda_acc} with respect to $\lambda$ as $H'_{\lambda}=h_0-\tfrac{\kappa_2}{\kappa_1\lambda^{\kappa_2+1}}e^{\kappa_3 Acc}$.
Taking $H'_{\lambda}=0$ yields the optimal scaling factor $\lambda^*=\sqrt[\kappa_2 + 1]{\tfrac{\kappa_2e^{\kappa_3 Acc}}{h_0\kappa_1}}$. We derive the following key insights regarding  $\lambda^*$.

\begin{remark}[Impact of Server Capability]
As the server's computational capacity increases ($h_0 \to 0$), the optimal scaling factor $\lambda^*$ grows. When the server possesses plenty of computational capacity, it can rapidly generate a vast synthetic images. As a result, a larger $\lambda$ accelerates FedFT without incurring a significant generation time.
\end{remark}

\begin{remark}[Impact of Target Accuracy]
The presence of $Acc$ in the exponent, $\lambda^* \propto e^{\tfrac{\kappa_3}{\kappa_2 + 1} Acc}$, reveals that the demand for synthetic data grows exponentially with the target accuracy. For low-accuracy requirements, the overhead of large-scale synthesis outweighs its benefits. However, for high-accuracy requirements, the FedFT time increases so sharply that a massive synthetic pre-training for model initialization becomes essential to reduce the total training time.
\end{remark}

\textbf{\large{\textcircled{\small{2}}}}  \textbf{Generating Synthetic Images and Pre-Training.}

After the optimal $\lambda^*$ is determined, the server generates a corresponding number of synthetic images (Alg. \ref{alg:FeDiSyn}, Lines \ref{alg1:cluster}-\ref{alg1:generation}) and pre-trains the LVM on the generated images dataset, then distributes LoRA-inserted model to devices (Alg. \ref{alg:FeDiSyn}, Lines \ref{alg1:pre-train}), as described in Section \ref{subsec:generation} and \ref{subsec:pretraining}.

\textbf{\large{\textcircled{\small{3}}}}  \textbf{Federated Fine-Tuning.}
The main idea of FedFT is to determine the LoRA configuration $\mathcal{X}^t$ to maximize the performance gain, while balancing completion times across devices through a bandwidth allocation strategy $\mathcal{B}^t$.
To handle device heterogeneity in FedFT, we design a contribution-aware LoRA configuration and bandwidth allocation (CLCBA) algorithm (Alg. \ref{alg:FeDiSyn}, Lines \ref{alg1:begin t}-\ref{alg1:end t}).
Since the local gradient $g_i(\widetilde{\mathbf{w}}_{i,l}^t)=\nabla f_{i}(\widetilde{\mathbf{w}}_{i,l}^{t})$ captures the contribution of $l$-th layer to the loss function \cite{sun2025exploring, li2026fedquad}, we use the L2 norm $G_i^t = \sum_{l \in L_i^t}||g_i(\widetilde{\mathbf{w}}_{i,l}^{t-1})||_2^2$ of the last round $t-1$ to quantify the contribution of $v_i$ at round $t$. The performance gain is defined as the sum of all devices' contributions divided by current round's duration
\begin{equation}
\label{eqn:reward}
    R^t\triangleq\tfrac{\sum_{v_i\in\mathcal{V}} G_i^t}{H^t}=\tfrac{\sum_{v_i\in\mathcal{V}}\sum_{l \in L_i^t}||g_i(\widetilde{\mathbf{w}}_{i,l}^{t-1})||_2^2}{H^t}.
\end{equation}

We first initialize the depths of all devices as 1, \ie, set $x_i^t = 1$ for all $v_i\in\mathcal{V}$ (Line \ref{alg1:set x_max}). In each iteration, we choose a device that obtains the maximum performance gain among all devices to add one LoRA layer. This process continues until the maximum performance gain is less than its value in the last iteration (Alg. \ref{alg:FeDiSyn}, Line \ref{alg1:begin while}-\ref{alg1:end while}). Meanwhile, in each iteration of the process of Alg. \ref{alg:FeDiSyn}, we call the water-filling based bandwidth allocation sub-algorithm (WFBA) to balance completion times across devices, which minimizes the completion time $H^t$ in Eq. \eqref{eqn:reward}.

\begin{algorithm}[t]
\caption{Water-Filling based Bandwidth Allocation (WFBA)}
\label{alg:WFBA}
\begin{algorithmic}[1]
\REQUIRE Bandwidth budget $B$, depth configuration $\mathcal{X}^{t}$, transmission power $p_i$, and channel gain $h_i^t$
\ENSURE Optimal bandwidth allocation $\mathcal{B}^t$
\FOR{$i \gets 1$ \TO $N$} \label{alg2:begin initial}
    \STATE Calculate $H_i^{t, \rm comp}\gets D_i(\overline{h}_i + x_i^t h_i)$
    \STATE $\gamma_i^t \gets \tau x_i^t/\log(1+\tfrac{p_i h_i^t}{\sigma^2})$
\ENDFOR \label{alg2:end initial}
\STATE $H_{\max}^{t, \rm comp} \gets \max_i H_i^{t, \rm comp}$\label{alg2:max time}
\STATE $q \gets 0$, $\mathcal{H}_q^t \gets 2H_{\max}^{t, \rm comp}$\label{alg2:begin newton}
\WHILE{$|\Delta H^t| > \zeta$}
    \STATE $\Phi(\mathcal{H}_{q}^t) = \sum_{i=1}^N \frac{\gamma_i^t}{\mathcal{H}_{q}^t - H_i^{t, \rm comp}}-B$
    \STATE $\mathcal{H}_{q+1}^t = \mathcal{H}_{q}^t - \frac{\Phi(\mathcal{H}_{q}^t)}{\Phi'(\mathcal{H}_{q}^t)}$
    \STATE $\Delta H^t \gets \mathcal{H}_{q+1}^t-\mathcal{H}_{q}^t$
    \STATE $q \gets q + 1$
\ENDWHILE   \label{alg2:end newton}
\FOR{$i \gets 1$ \TO $N$} \label{alg2:begin return}
    \STATE $B_i^t \gets \gamma_i^t/(\mathcal{H}_{q}^t-H_i^{t, \rm comp})$
\ENDFOR
\RETURN $\mathcal{B}^t$ \label{alg2:end return}
\end{algorithmic}
\end{algorithm}

\begin{figure}[t]
\centering
\includegraphics[width=\linewidth]{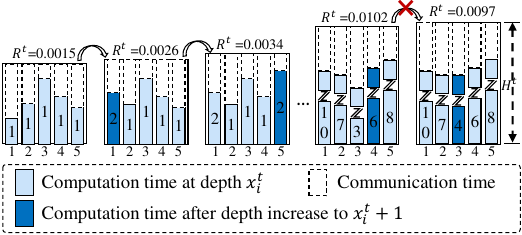}
\vspace{-7mm}
\caption{An example of CLCBA.
\textmd{Initially, devices $v_1-v_{5}$ have $x_i^t=1$.
In the first iteration, increasing the depth of $v_1$ yields $R^t=$ 0.0026, and further increasing $v_5$'s depth raises $R^t$ to 0.0034. This continues until the maximum $R^t=$ 0.0102. Further increasing the depth of other devices will decrease $R^t$ to 0.0097, so CLCBA terminates and returns the LoRA configuration $\mathcal{X}^t$.
In this process, WFBA is invoked to obtain $\mathcal{B}^t$, balancing the completion time of all devices.}}
\label{fig:alg_fic}
\vspace{-4mm}
\end{figure}

The WFBA sub-algorithm is shown in Alg. \ref{alg:WFBA}. First, WFBA calculates the maximum computation time $H_{\max}^{t, \rm comp}$ of all devices (Alg. \ref{alg:WFBA}, Lines~\ref{alg2:begin initial}-\ref{alg2:max time}). Since the waterline must exceed the maximum computation time $H_{\max}^{t, \rm comp}$, WFBA initializes the waterline as $\mathcal{H}_0^t = 2 H_{\max}^{t, \rm comp}$ and iteratively solve its optimal value using Newton's method (Alg. \ref{alg:WFBA}, Lines~\ref{alg2:begin newton}-\ref{alg2:end newton}). Finally, WFBA calculates the optimal bandwidth allocation $B_i^t$ for each device $v_i$ based on the converged waterline and returns the bandwidth configuration $\mathcal{B}^t$ (Alg. \ref{alg:WFBA}, Lines \ref{alg2:begin return}-\ref{alg2:end return}). The intuitive diagram of CLCBA to obtain $\mathcal{X}^t$ and $\mathcal{B}^t$ are shown in Fig. \ref{fig:alg_fic}.

\section{Performance Evaluation}\label{sec:evaluation}
\subsection{System Setup}\label{subsec:setup}

\begin{figure}[h]
    \centering
    \includegraphics[width=0.8\linewidth]{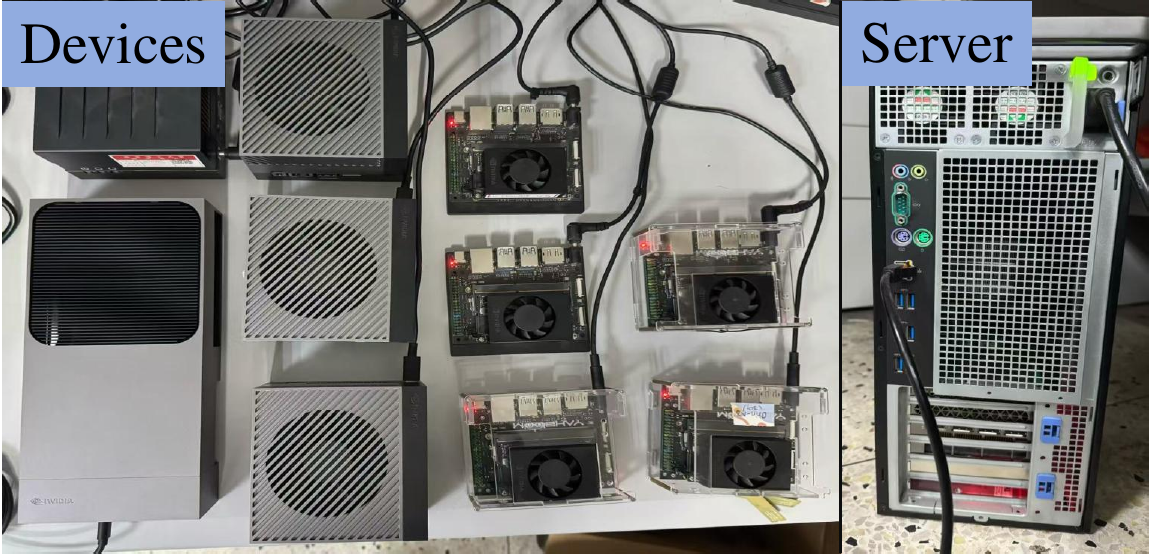}
    \vspace{-2mm}
    \caption{Hardware devices of our testbed.}
    \label{fig:testbed_fig}
    \vspace{-5mm}
\end{figure}

\textbf{Environment Settings.}
We implement a FedFT experimental testbed comprising one server and 10 heterogeneous devices, as shown in Fig. \ref{fig:testbed_fig}. The server is equipped with an NVIDIA GeForce RTX 4090 GPU with 48GB GDDR6X. We use 1 Jetson AGX Xavier kit, 5 Jetson Orin NX kits, 3 Jetson AGX Orin kits, and 1 Jetson AGX Thor kit as heterogeneous devices. Particularly, we simulate more practical wireless communication in an edge environment instead of high-bandwidth laboratory networks, where communication time is typically negligible compared to computation time. Similar to \cite{wang2025federated}, the server is at (0, 0), and each device is uniformly distributed within a circular area of 50 meters radius centered at coordinates $(300,0)$. The large-scale fading is represented as $\left( d/d_{0} \right)^{-3.5}$, where $d_{0} = 10$ meters is the reference distance. The noise power is set as $\sigma^2=10^{-11}$W, with a total system bandwidth of $B = 10$MHz.

\textbf{Models and Datasets.}
We train ViT-B/16 \cite{dosovitskiy2020image} on CIFAR-10 \cite{krizhevsky2009learning} and Caltech-101 \cite{li2004Learning} datasets, and train ViT-L/16 \cite{dosovitskiy2020image} on Food-101 \cite{bossard14food} dataset. ViT-B/16 and ViT-L/16 consist of 86 and 307 million parameters, respectively.

\begin{itemize}[leftmargin=*]
    \item \textbf{CIFAR-10 \cite{krizhevsky2009learning}} is composed of 60,000 color images, which are divided into 10 classes, each containing 6,000 images. The dataset is further split into 50,000 training images and 10,000 test images.
    \item \textbf{Caltech-101 \cite{li2004Learning}} comprises 101 classes objects, totaling 9,144 samples. The number of samples per category ranges from 40 to 800 images, with each category averaging approximately 50 samples. We divide it into a 4:1 training-test split ratio.
    \item \textbf{Food-101 \cite{bossard14food}} contains a series of food datasets divided into 101 classes with high inter-class similarity. It comprises 101,000 samples, with each class including 750 training and 250 test samples.
\end{itemize}

We analyze FedFT performance under both IID and non-IID data among devices. Similar to \cite{shi2026dystop}, for the non-IID data case, we adopt the Dirichlet distribution \cite{tzu2019measuring} with $\alpha_{\rm dir}=0.3$ to partition the dataset. In addition, we deploy the LLaMa-7B \cite{touvron2023llama} to enrich prompts and Stable-Diffusion-v1-5 \cite{rombach2022high} to generate synthetic images as described in Section \ref{subsec:generation}.

\textbf{Benchmarks.} We compare FeDiSyn against two solutions with synthetic image pre-training (GPT-FL \cite{zhang2025gpt}, FGL \cite{zhang2023federated}) and two solutions deploying LoRA-based FedFT (SFLF \cite{wang2025federated}, CAFF \cite{pfeiffer2024efficient}). Specifically, GPT-FL and FGL generate images based on uploaded labels or text embeddings from devices, while SFLF and CAFF determine each device's LoRA depth based on its capacity. For a fair comparison, FeDiSyn, SFLF, and CAFF adopt the same LoRA rank.

\begin{itemize}[leftmargin=*]
    \item \textbf{GPT-FL\cite{zhang2025gpt}} is a synthetic image pre-training based approach that expands device-uploaded labels into prompts and generates images for pre-training LVM, while using FedAvg \cite{mcmahan2017communication} for FedFT. 
    \item \textbf{FGL\cite{zhang2023federated}} is a synthetic image pre-training based approach where devices use BLIP to generate prompts for local images, and the server employs these prompts to generate images for pre-training LVM, while utilizing FedAvg for FedFT. 
    \item \textbf{SFLF\cite{wang2025federated}} is a LoRA-based approach for FedFT, where the LVM is initialized with ImageNet parameters, and all devices fine-tune all transformer layers of the LVM.
    \item \textbf{CAFF\cite{pfeiffer2024efficient}} is an advanced LoRA-based approach that initializes LVM with ImageNet parameters, and assigns appropriate LoRA depths to each device to address system heterogeneity. 
\end{itemize}

\subsection{Evaluation Results}\label{subsec:result}

\begin{figure}[t]
\includegraphics[width=0.325\linewidth]{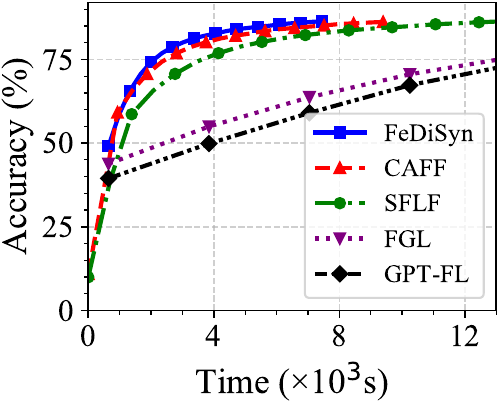}
\includegraphics[width=0.325\linewidth]{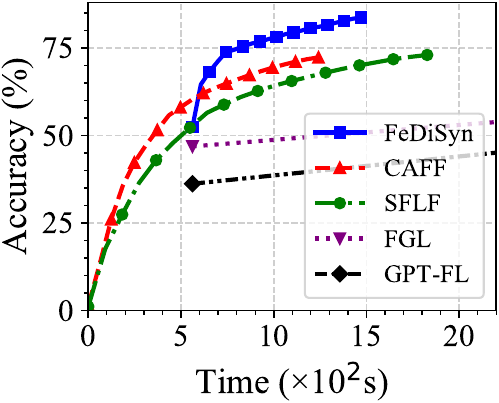}
\includegraphics[width=0.325\linewidth]{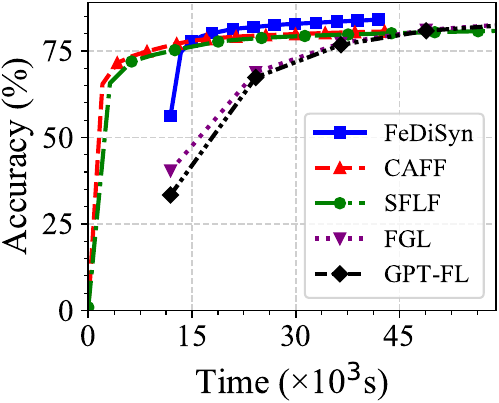}
\vspace{-3mm}
\small{\caption{The test accuracy over different datasets (IID). \textit{Left}: CIFAR-10; \textit{Middle}: Caltech-101; \textit{Right}: Food-101.}\label{fig:iid_acc}}
\vspace{-5mm}
\end{figure}

\begin{figure}[t]
\includegraphics[width=0.325\linewidth]{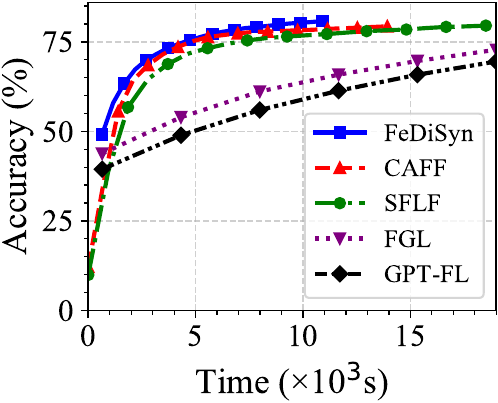}
\includegraphics[width=0.325\linewidth]{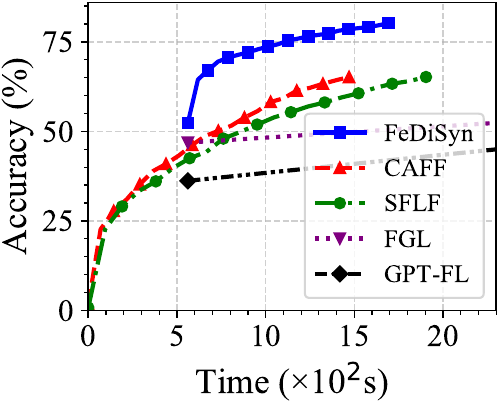}
\includegraphics[width=0.325\linewidth]{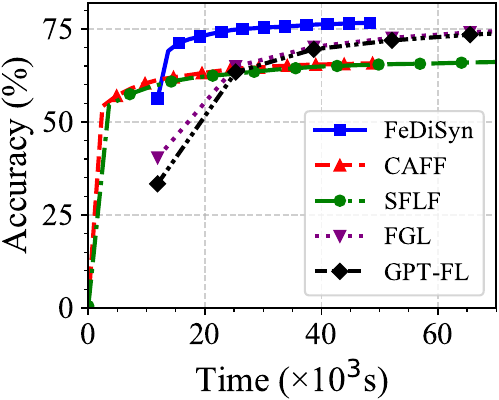}
\vspace{-3mm}
\small{\caption{The test accuracy over different datasets (non-IID). \textit{Left}: CIFAR-10; \textit{Middle}: Caltech-101; \textit{Right}: Food-101.}\label{fig:noniid_acc}}
\vspace{-5mm}
\end{figure}

\begin{figure}[t]
\includegraphics[width=0.325\linewidth]{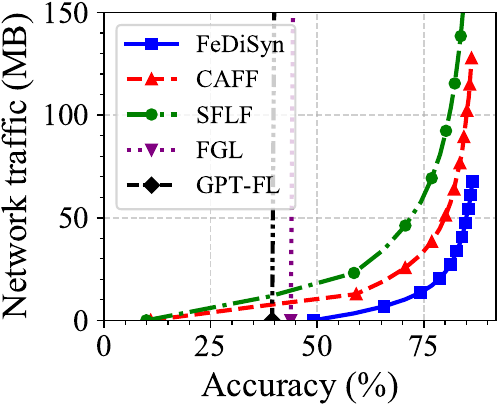}
\includegraphics[width=0.325\linewidth]{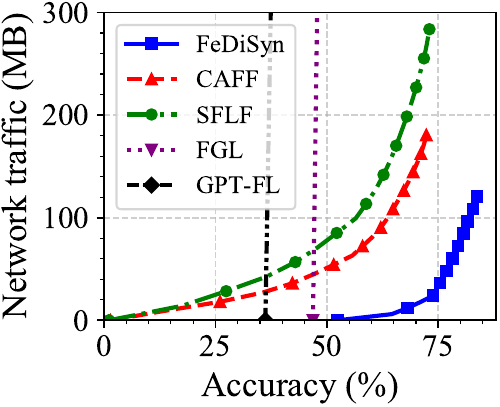}
\includegraphics[width=0.325\linewidth]{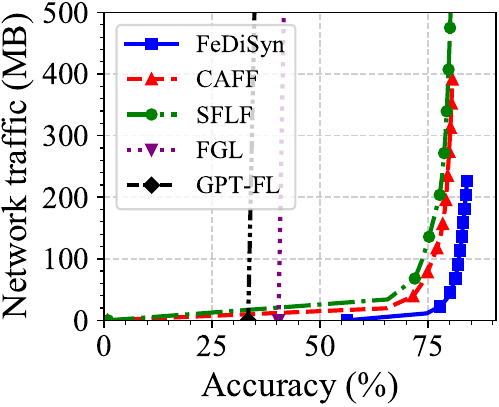}
\vspace{-3mm}
\small{\caption{The network traffic over different datasets (IID). \textit{Left}: CIFAR-10; \textit{Middle}: Caltech-101; \textit{Right}: Food-101.}\label{fig:iid_bandwidth}}
\vspace{-5mm}
\end{figure}

\begin{figure}[t]
\includegraphics[width=0.325\linewidth]{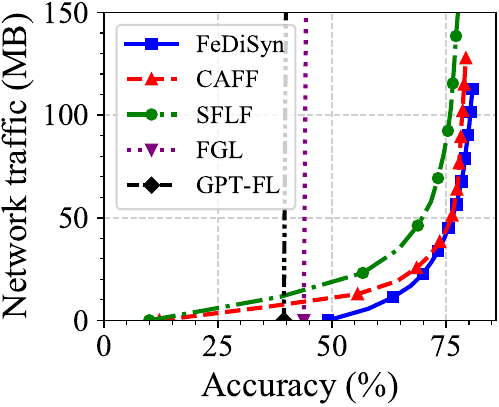}
\includegraphics[width=0.325\linewidth]{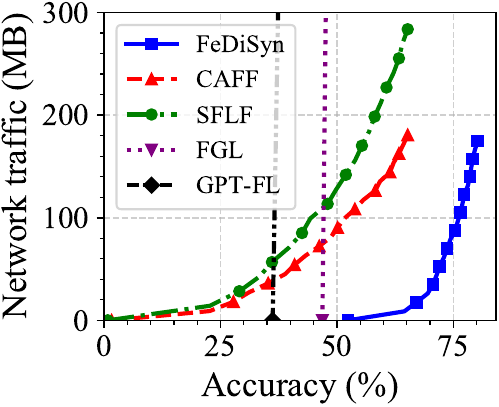}
\includegraphics[width=0.325\linewidth]{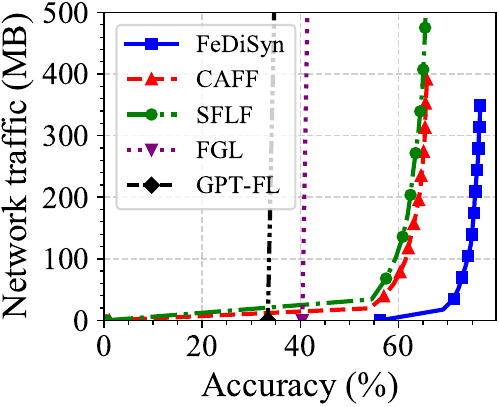}
\vspace{-3mm}
\small{\caption{The network traffic over different datasets (non-IID). \textit{Left}: CIFAR-10; \textit{Middle}: Caltech-101; \textit{Right}: Food-101.}\label{fig:noniid_bandwidth}}
\vspace{-5mm}
\end{figure}

\textbf{Training Performance.}
We evaluate the FedFT process of our FeDiSyn method and the benchmarks on the CIFAR-10, Caltech-101 and Food-101 datasets under both IID and non-IID data distributions, as illustrated in Figs. \ref{fig:iid_acc} and \ref{fig:noniid_acc}.
Overall, FeDiSyn consistently achieves the fastest convergence and highest accuracy on all datasets under both IID and non-IID data. For example, as shown in the third plot of Fig. \ref{fig:iid_acc}, under IID data on Food-101, FeDiSyn reaches 80\% accuracy in 17,584s, while CAFF, SFLF, FGL, and GPT-FL require 31,905s, 41,885s, 44,783s, and 45,383s, respectively, meaning FeDiSyn
reduces 44.9\%, 58.0\%, 60.7\%, and 61.3\% completion time compared to these benchmarks. Moreover, under non-IID data, FeDiSyn's advantage becomes even more pronounced compared to other solutions. As shown in the third plot of Fig. \ref{fig:noniid_acc}, under non-IID data, FeDiSyn reduces 62.0\%, 70.5\%, 52.5\%, and 56.2\% time for reaching 65\% accuracy compared to CAFF, SFLF, FGL, and GPT-FL, respectively. This is because FeDiSyn pre-training over synthetic images that enhances the generalization ability of initial model, alleviating the impact of non-IID data \cite{jhunjhunwala2025initialization}.

\textbf{Network Traffic Consumption.}
In Fig. \ref{fig:iid_bandwidth} and \ref{fig:noniid_bandwidth}, we compare the network traffic consumption by different solutions to achieve certain accuracies on CIFAR-10 Caltech-101 and Food-101, respectively. By the results, FeDiSyn significantly reduces the network traffic compared with the benchmarks.
For example, under non-IID data on Food-101, the consumed network traffic of FeDiSyn, CAFF, SFLF, FGL, and GPT-FL is 7.4MB, 262.1MB, 399.5MB, 12,002.8MB, and 13,676.3MB, respectively. FeDiSyn can reduce network traffic by 97.2\%, 98.1\%, 99.9\%, and 99.9\% compared with these methods. Compared with the LoRA-based methods CAFF and SFLF, FeDiSyn adopts a more efficient LoRA configuration strategy, thereby reducing network traffic when achieving the target accuracy. In addition, FGL and GPT-FL rely on full parameter FedFT, which incurs extremely high communication overhead.

The complete experimental results of FedFT performance and network traffic consumption are provided in Table \ref{tbl:overall_test}.

\begin{table}[t]
\centering
\footnotesize
\caption{The fine-tuning time ($10^2$s) and network traffic (MB) to reach target accuracy.}
\vspace{-3mm}
\label{tbl:overall_test}
\begin{tabular}{c|c|cccc>{\columncolor{gray!20}}c}
\toprule
Dataset & Metric & GPT-FL & FGL & SFLF & CAFF & \textbf{FeDiSyn} \\
\midrule

\multirow{2}{*}{\makecell{CIFAR-10\\ (IID, 85\%)}}
    & Time    & 232.5 & 220.3 & 103.5 & 73.3 & \textbf{55.1} \\
    & Traffic & 23243.5 & 21798.8 & 172.7 & 99.8 & \textbf{50.3} \\
\midrule
\multirow{2}{*}{\makecell{CIFAR-10\\ (Non-IID, 80\%)}}
    & Time    & 348.6 & 341.3 & 203.7 & 161.1 & \textbf{92.3} \\
    & Traffic & 31051.5 & 29840.5 & 254.0 & 146.0 & \textbf{93.5} \\
\midrule
\multirow{2}{*}{\makecell{Caltech-101\\ (IID, 70\%)}}
    & Time    & 113.7 & 96.4 & 14.6 & 10.3 & \textbf{6.8} \\
    & Traffic & 14547.6 & 12198.6 & 226.2 & 150.1 & \textbf{15.7} \\
\midrule
\multirow{2}{*}{\makecell{Caltech-101\\ (Non-IID, 65\%)}}
    & Time    & 147.5 & 116.7 & 18.9 & 14.5 & \textbf{6.3} \\
    & Traffic & 18564.2 & 14755.1 & 279.5 & 177.8 & \textbf{10.1} \\
\midrule
\multirow{2}{*}{\makecell{Food-101\\ (IID, 80\%)}}
    & Time    & 453.8 & 447.8 & 418.9 & 319.1 & \textbf{175.8} \\
    & Traffic & 30763.4 & 30375.9 & 457.3 & 290.1 & \textbf{42.5} \\
\midrule
\multirow{2}{*}{\makecell{Food-101\\ (Non-IID, 65\%)}}
    & Time    & 284.9 & 262.2 & 423.1 & 328.4 & \textbf{124.7} \\
    & Traffic & 13676.3 & 12002.8 & 399.5 & 262.1 & \textbf{7.4} \\
\bottomrule
\end{tabular}
\vspace{-5mm}
\end{table}

\begin{figure}[t]
\includegraphics[width=0.325\linewidth]{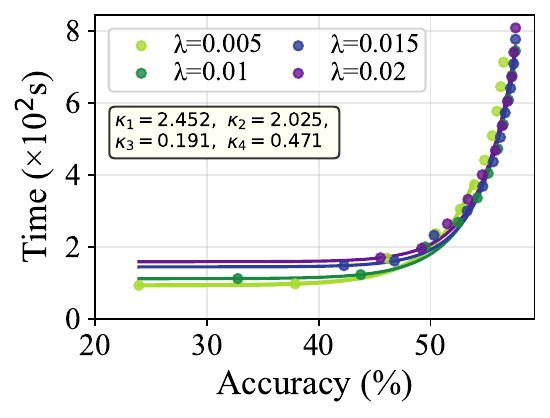}
\includegraphics[width=0.325\linewidth]{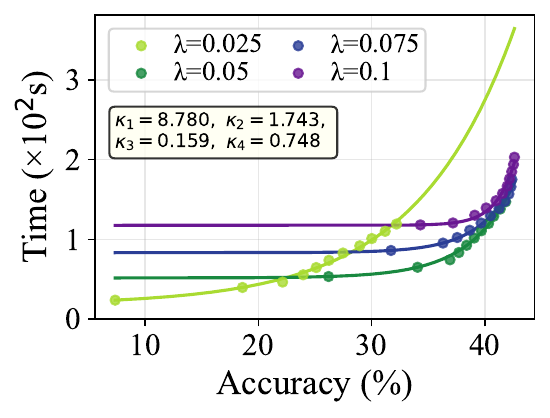}
\includegraphics[width=0.325\linewidth]{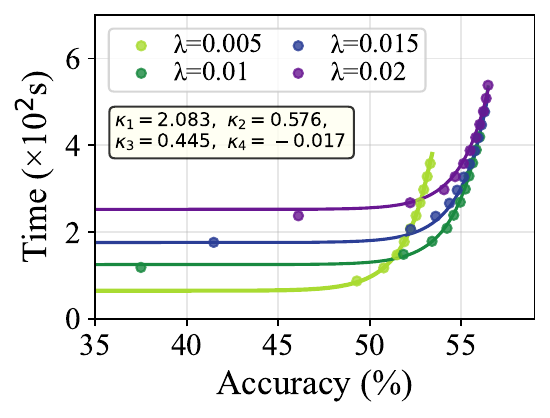}
\vspace{-3mm}
\small{\caption{Fitting Eq. \eqref{eqn:time_lambda_acc} with small $\lambda$. \textit{Left}: CIFAR-10; \textit{Middle}: Caltech-101; \textit{Right}: Food-101.}\label{fig:fitting}}
\vspace{-5mm}
\end{figure}

\begin{figure}[t]
\includegraphics[width=0.325\linewidth]{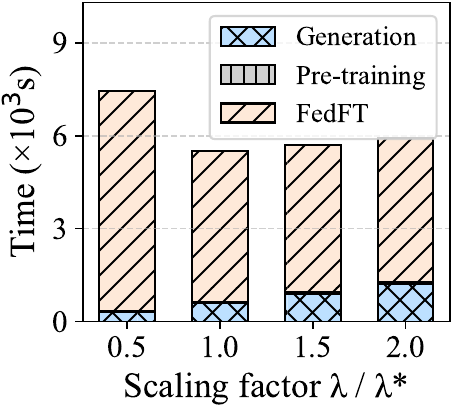}
\includegraphics[width=0.325\linewidth]{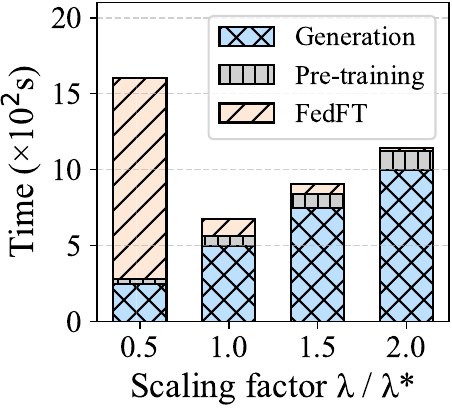}
\includegraphics[width=0.325\linewidth]{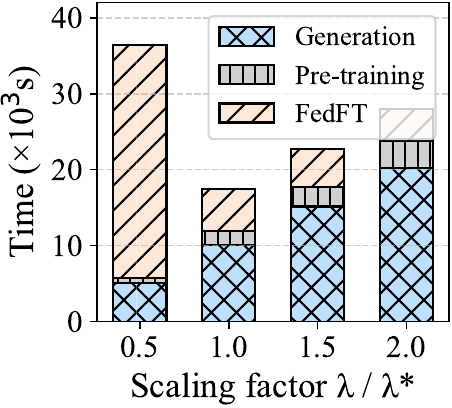}
\vspace{-3mm}
\small{\caption{The time to reach target accuracy under different scaling factor $\lambda$. \textit{Left}: CIFAR-10 (85\%); \textit{Middle}: Caltech-101 (70\%); \textit{Right}: Food-101 (75\%).}\label{fig:lambda_total_time}}
\vspace{-5mm}
\end{figure}

\textbf{Impact of Synthetic Image Volume.} To estimate the values of coefficients $\kappa_1-\kappa_4$ in Eq. \eqref{eqn:time_lambda_acc}, we fit this equation in a short calibration phase using a set
of small values of $\lambda$ via nonlinear least-squares (\eg, Levenberg-Marquardt method). Fig. \ref{fig:fitting} shows that Eq. \eqref{eqn:time_lambda_acc} well describes the relationship among total training time, accuracy and the volume of generated data. Then we obtain the optimal $\lambda^*=\sqrt[\kappa_2 + 1]{\tfrac{\kappa_2e^{\kappa_3 Acc}}{h_0\kappa_1}}$ by the estimated coefficients.

Fig. \ref{fig:lambda_total_time} shows the time to reach target accuracy under different scaling factor  $\lambda=0.5\lambda^*, \lambda^*, 1.5\lambda^*$ and $2\lambda^*$ on CIFAR-10, Caltech-101 and Food-101. $Acc$ is set as 85\%, 70\% and 80\% for CIFAR-10, Caltech-101 and Food-101, respectively. We observe that the generation and pre-training time increase linearly with $\lambda$. For example, for Food-101, the generation times are 5,050s, 10,100s, 15,150s, and 20,200s for $\lambda=0.5\lambda^*, \lambda^*, 1.5\lambda^*$, and $2\lambda^*$, respectively. However, when the value of $\lambda$ is too large or too small, the total completion time to achieve the target accuracy will be relatively long. For example, for Food-101, when $\lambda$ increases from $0.5\lambda^*$ to $2\lambda^*$, the time required to achieve 80\% accuracy is 36,351.0s, 17,583.9s, 22,671.1s, and 27,943.9s, respectively. This is because insufficiently generated images weaken pre-training effectiveness for FedFT, whereas excessive images introduce substantial computational overhead, thereby prolonging image generation. The experimental results in Figs. \ref{fig:fitting}-\ref{fig:lambda_total_time} show that the values of $\lambda^*$ obtained by the empirical model have the shortest completion time compared to other values.

\begin{figure}[t]
\includegraphics[width=0.325\linewidth]{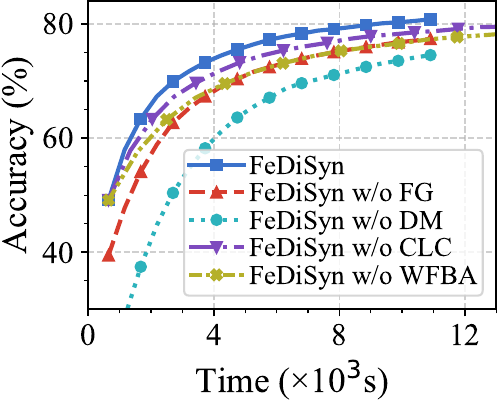}
\includegraphics[width=0.325\linewidth]{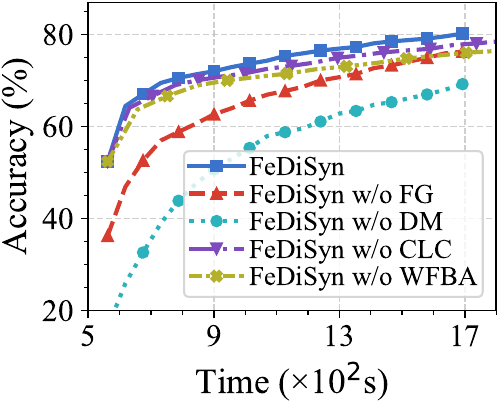}
\includegraphics[width=0.325\linewidth]{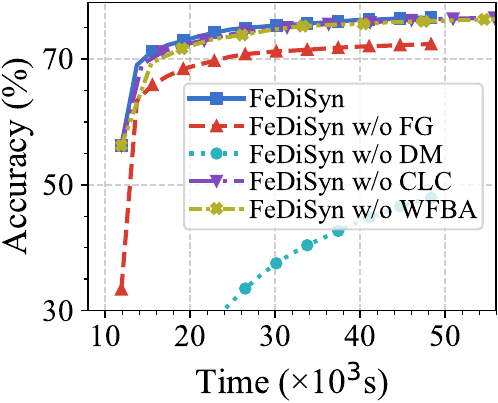}
\vspace{-3mm}
\small{\caption{Ablation study on accuracy. \textit{Left}: CIFAR-10; \textit{Middle}: Caltech-101; \textit{Right}: Food-101.}\label{fig:ablation}}
\vspace{-5mm}
\end{figure}

\begin{figure}[t]
\includegraphics[width=0.325\linewidth]{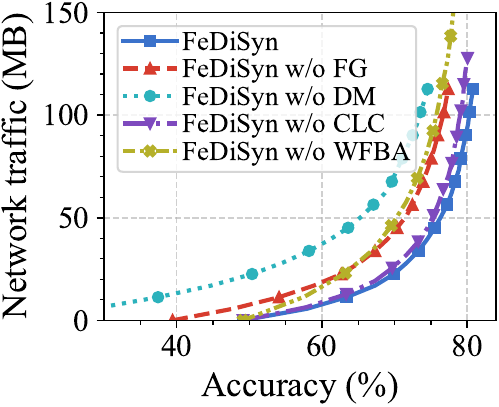}
\includegraphics[width=0.325\linewidth]{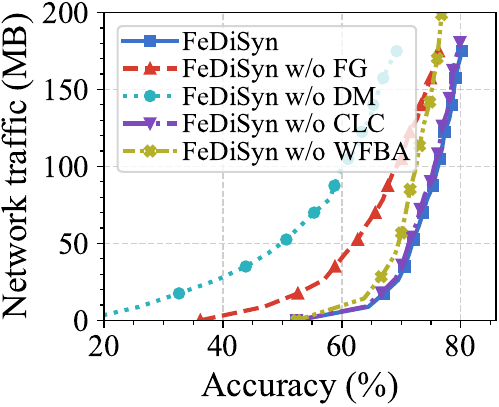}
\includegraphics[width=0.325\linewidth]{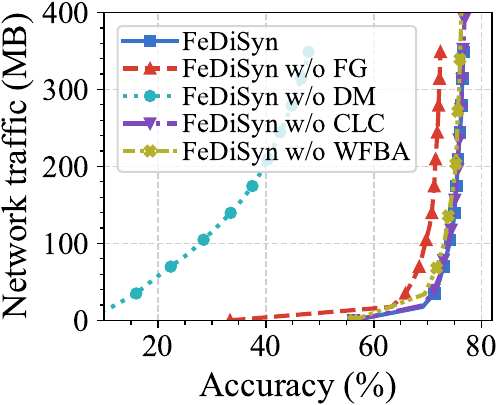}
\vspace{-3mm}
\small{\caption{Ablation study on Bandwidth. \textit{Left}: CIFAR-10; \textit{Middle}: Caltech-101; \textit{Right}: Food-101.}\label{fig:ablation_bandwidth}}
\vspace{-5mm}
\end{figure}

\textbf{Ablation Study.} FeDiSyn comprises two primary modules: 1) a feature-guided (FG) synthetic generation module with a diffusion model (DM), 2) a LoRA-based FedFT module deploying contribution-aware LoRA configuration (CLC) and water-filling bandwidth allocation (WFBA). We conduct fine-grained ablation experiments on CIFAR-10, Caltech-101, and Food-101 in Figs. \ref{fig:ablation} and \ref{fig:ablation_bandwidth} to evaluate the individual contribution of each sub-module. FeDiSyn consistently outperforms all variants, including those without FG, CLC, WFBA, or those utilizing GANs instead of DMs. For example, on CIFAR-10, FeDiSyn achieves 74\% accuracy with 42.4\%, 61.4\%, 22.8\%, and 41.7\% time reduction and reduces network traffic usage by 47.0\%, 65.6\%, 12.5\%, and 51.7\%, compared to FeDiSyn w/o FG, FeDiSyn w/o DM, FeDiSyn w/o CLC, and FeDiSyn w/o WFBA, respectively. We also observe that the relative importance of synthetic pre-training grows with model scale. For example, for ViT-B/16 training on CIFAR-10 and Caltech-101, the performance gaps among different variants are relatively small, but for ViT-L/16 training on Food-101, FeDiSyn w/o CLC and w/o WFBA significantly outperform FeDiSyn w/o FG and w/o DM, even approaching the performance of FeDiSyn. This is because larger models possess a higher degree of representational complexity, making a high-quality pre-training for initialization far more critical than fine-tuning stage.

\section{Conclusion}\label{sec:conclusion}

In this paper, we propose FeDiSyn, a unified framework that holistically considers both pre-training and federated fine-tuning (FedFT) of large vision models to address resource constraints, system heterogeneity, and non-IID data. FeDiSyn introduces a scaling law for synthetic pre-training under FedFT setting to determine the optimal number of synthetic images. A diffusion model is deployed on the server to generate synthetic images that align with the device's local image label and feature distributions, thereby efficiently pre-training a superior initial global model for FedFT. During FedFT, we design a contribution-aware LoRA configuration and bandwidth allocation algorithm to enhance FedFT performance while addressing system heterogeneity. Experimental results indicate the superiority of our proposed framework compared with state-of-the-art solutions.

\begin{acks}
We thank the anonymous reviewers for their valuable comments. This work was supported in part by the National Natural Science Foundation of China (NSFC) under Grants 62402537, 62572244, 62402226 and U25A20442, in part by the Natural Science Foundation of Jiangsu Province under Grant BK20241453, and in part by the Fundamental Research Funds for the Central Universities under Grant 30925010408. The corresponding author of this paper is Junlong Zhou.
\end{acks}



\bibliographystyle{ACM-Reference-Format}
\bibliography{contents/refs}

@article{krizhevsky2012imagenet,
  title={Imagenet classification with deep convolutional neural networks},
  author={Krizhevsky, Alex and others},
  journal={Advances in neural information processing systems},
  volume={25},
  year={2012}
}

@article{zhu2020toward,
  title={Toward an intelligent edge: Wireless communication meets machine learning},
  author={Zhu, Guangxu and Liu, Dongzhu and Du, Yuqing and You, Changsheng and Zhang, Jun and Huang, Kaibin},
  journal={IEEE communications magazine},
  volume={58},
  number={1},
  pages={19--25},
  year={2020},
  publisher={IEEE}
}

@article{zhou2022swarm,
  title={Swarm intelligence-based task scheduling for enhancing security for IoT devices},
  author={Zhou, Junlong and Shen, Yufan and Li, Liying and Zhuo, Cheng and Chen, Mingsong},
  journal={IEEE Transactions on Computer-Aided Design of Integrated Circuits and Systems},
  volume={42},
  number={6},
  pages={1756--1769},
  year={2022},
  publisher={IEEE}
}

@inproceedings{yang2024exploring,
  title={Exploring one-shot semi-supervised federated learning with pre-trained diffusion models},
  author={Yang, Mingzhao and Su, Shangchao and Li, Bin and Xue, Xiangyang},
  booktitle={Proceedings of the AAAI conference on artificial intelligence},
  volume={38},
  number={15},
  pages={16325--16333},
  year={2024}
}

@inproceedings{deng2009imagenet,
  title={Imagenet: A large-scale hierarchical image database},
  author={Deng, Jia and Dong, Wei and Socher, Richard and Li, Li-Jia and Li, Kai and Fei-Fei, Li},
  booktitle={2009 IEEE conference on computer vision and pattern recognition},
  pages={248--255},
  year={2009}
}

@article{wang2019adaptive,
  author={Wang, Shiqiang and Tuor, Tiffany and Salonidis, Theodoros and Leung, Kin K and Makaya, Christian and He, Ting and Chan, Kevin},
  journal={IEEE Journal on Selected Areas in Communications},
  title={Adaptive Federated Learning in Resource Constrained Edge Computing Systems},
  year={2019},
  volume={37},
  number={6},
  pages={1205-1221}
}

@article{jhunjhunwala2025initialization,
  title={Initialization Matters: Unraveling the Impact of Pre-Training on Federated Learning},
  author={Jhunjhunwala, Divyansh and Sharma, Pranay and Xu, Zheng and Joshi, Gauri},
  journal={Transactions on machine learning research},
  year={2025}
}

@article{ma2025asynchronous,
  title={Asynchronous Federated Learning Over Non-IID Data via Over-the-Air Computation},
  author={Ma, Qianpiao and Song, Xiaozhu and Zhou, Junlong and Wang, Haibo and Liao, Yunming and Liu, Jianchun and Xu, Hongli},
  journal={IEEE Transactions on Networking},
  year={2026},
  volume={34},
  number={},
  pages={2165-2180},
  publisher={IEEE}
}

@article{zhou2024federated,
  author={Zhou, Tianfei and Yuan, Ye and Wang, Binglu and Konukoglu, Ender},
  journal={IEEE Transactions on Pattern Analysis and Machine Intelligence},
  title={Federated Feature Augmentation and Alignment},
  year={2024},
  volume={46},
  number={12},
  pages={11119-11135}
}

@inproceedings{yan2025simple,
  title={A simple data augmentation for feature distribution skewed federated learning},
  author={Yan, Yunlu and Fu, Huazhu and Li, Yuexiang and Xie, Jinheng and Ma, Jun and Yang, Guang and Zhu, Lei},
  booktitle={Proceedings of the IEEE/CVF Conference on Computer Vision and Pattern Recognition},
  pages={25749--25758},
  year={2025}
}

@article{zhang2023federated,
  title={Federated generative learning with foundation models},
  author={Zhang, Jie and Qi, Xiaohua and Zhao, Bo},
  journal={arXiv preprint arXiv:2306.16064},
  year={2023}
}

@inproceedings{zhang2023fedpetuning,
  title={FedPETuning: When federated learning meets the parameter-efficient tuning methods of pre-trained language models},
  author={Zhang, Zhuo and Yang, Yuanhang and Dai, Yong and Wang, Qifan and Yu, Yue and Qu, Lizhen and Xu, Zenglin},
  booktitle={Association for Computational Linguistics},
  pages={9963--9977},
  year={2023}
}

@inproceedings{zhang2025gpt,
  title={Gpt-fl: Generative pre-trained model-assisted federated learning},
  author={Zhang, Tuo and Feng, Tiantian and Alam, Samiul and Dimitriadis, Dimitrios and Lee, Sunwoo and Zhang, Mi and Narayanan, Shrikanth S and Avestimehr, Salman},
  booktitle={Proceedings of the Computer Vision and Pattern Recognition Conference},
  pages={1761--1770},
  year={2025}
}

@book{krizhevsky2009learning,
  title={Learning multiple layers of features from tiny images},
  author={Krizhevsky, Alex and Hinton, Geoffrey and others},
  year={2009},
  publisher={Citeseer}
}

@article{ma2021fedsa,
  author={Ma, Qianpiao and Xu, Yang and Xu, Hongli and Jiang, Zhida and Huang, Liusheng and Huang, He},
  journal={IEEE Journal on Selected Areas in Communications},
  title={FedSA: A Semi-Asynchronous Federated Learning Mechanism in Heterogeneous Edge Computing},
  year={2021},
  volume={39},
  number={12},
  pages={3654-3672}
}

@inproceedings{ma2025air,
  title={Air-FedGA: A Grouping Asynchronous Federated Learning Mechanism Exploiting Over-The-Air Computation},
  author={Ma, Qianpiao and Zhou, Junlong and Hou, Xiangpeng and Liu, Jianchun and Xu, Hongli and Miao, Jianeng and Jia, Qingmin},
  booktitle={2025 IEEE International Parallel and Distributed Processing Symposium (IPDPS)},
  pages={1--12},
  year={2025},
  organization={IEEE}
}

@article{he2022synthetic,
  title={Is synthetic data from generative models ready for image recognition?},
  author={He, Ruifei and Sun, Shuyang and Yu, Xin and Xue, Chuhui and Zhang, Wenqing and Torr, Philip and Bai, Song and Qi, Xiaojuan},
  journal={International Conference on Learning Representations},
  year={2023}
}

@inproceedings{rombach2022high,
  title={High-resolution image synthesis with latent diffusion models},
  author={Rombach, Robin and Blattmann, Andreas and Lorenz, Dominik and Esser, Patrick and Ommer, Bj{\"o}rn},
  booktitle={Proceedings of the IEEE/CVF Conference on Computer Vision and Pattern Recognition},
  pages={10684--10695},
  year={2022}
}

@inproceedings{liu2022pseudo,
  title={Pseudo Numerical Methods for Diffusion Models on Manifolds},
  author={Liu, Luping and Ren, Yi and Lin, Zhijie and Zhao, Zhou},
  booktitle={International Conference on Learning Representations},
  year={2022}
}

@article{hu2022lora,
  title={Lora: Low-rank adaptation of large language models.},
  author={Hu, Edward J and Shen, Yelong and Wallis, Phillip and Allen-Zhu, Zeyuan and Li, Yuanzhi and Wang, Shean and Wang, Lu and Chen, Weizhu and others},
  journal={International Conference on Learning Representations},
  year={2022}
}

@InProceedings{radford2021learning,
  title = {Learning Transferable Visual Models From Natural Language Supervision},
  author={Radford, Alec and Kim, Jong Wook and Hallacy, Chris and Ramesh, Aditya and Goh, Gabriel and Agarwal, Sandhini and Sastry, Girish and Askell, Amanda and Mishkin, Pamela and Clark, Jack and others},
  booktitle = {Proceedings of the 38th International Conference on Machine Learning},
  pages = {8748--8763},
  year = {2021},
  volume = {139}
}

@InProceedings{mcmahan2017communication,
  title = {Communication-Efficient Learning of Deep Networks from Decentralized Data},
 author={McMahan, Brendan and Moore, Eider and Ramage, Daniel and Hampson, Seth and y Arcas, Blaise Aguera},
  booktitle = {Proceedings of Machine Learning Research},
  pages = {1273--1282},
  year = {2017},
  volume = {54}
}

@inproceedings{sohl2015deep,
  title={Deep unsupervised learning using nonequilibrium thermodynamics},
  author={Sohl-Dickstein, Jascha and Weiss, Eric and Maheswaranathan, Niru and Ganguli, Surya},
  booktitle={International Conference on Machine Learning},
  pages={2256--2265},
  year={2015}
}

@ARTICLE{ma2024feduc,
  author={Ma, Qianpiao and Xu, Yang and Xu, Hongli and Liu, Jianchun and Huang, Liusheng},
  journal={IEEE Transactions on Mobile Computing},
  title={FedUC: A Unified Clustering Approach for Hierarchical Federated Learning},
  year={2024},
  volume={23},
  number={10},
  pages={9737-9756}
}

@inproceedings{zhang2023adaptive,
    title={Adaptive Budget Allocation for Parameter-Efficient Fine-Tuning },
    author={Qingru Zhang and Minshuo Chen and Alexander Bukharin and Pengcheng He and Yu Cheng and Weizhu Chen and Tuo Zhao},
    booktitle={The Eleventh International Conference on Learning Representations},
    year={2023}
}

@article{liu2025adaptive,
  title={Adaptive parameter-efficient federated fine-tuning on heterogeneous devices},
  author={Liu, Jun and Liao, Yunming and Xu, Hongli and Xu, Yang and Liu, Jianchun and Qian, Chen},
  journal={IEEE Transactions on Mobile Computing},
  year={2025}
}

@inproceedings{dosovitskiy2020image,
  title={An Image is Worth 16x16 Words: Transformers for Image Recognition at Scale},
  author={Dosovitskiy, Alexey and Beyer, Lucas and Kolesnikov, Alexander and Weissenborn, Dirk and Zhai, Xiaohua and Unterthiner, Thomas and  Dehghani, Mostafa and Minderer, Matthias and Heigold, Georg and Gelly, Sylvain and Uszkoreit, Jakob and Houlsby, Neil},
  journal={International Conference on Learning Representations},
  year={2021}
}

@inproceedings{su2024fedra,
  title={Fedra: A random allocation strategy for federated tuning to unleash the power of heterogeneous clients},
  author={Su, Shangchao and Li, Bin and Xue, Xiangyang},
  booktitle={European Conference on Computer Vision},
  pages={342--358},
  year={2024},
  organization={Springer}
}

@inproceedings{xie2024perada,
  title={Perada: Parameter-efficient federated learning personalization with generalization guarantees},
  author={Xie, Chulin and Huang, De-An and Chu, Wenda and Xu, Daguang and Xiao, Chaowei and Li, Bo and Anandkumar, Anima},
  booktitle={Proceedings of the IEEE/CVF Conference on Computer Vision and Pattern Recognition},
  pages={23838--23848},
  year={2024}
}

@ARTICLE{guo2024explore,
  author={Guo, Tao and Guo, Song and Wang, Junxiao},
  journal={IEEE Transactions on Mobile Computing},
  title={Explore and Cure: Unveiling Sample Effectiveness With Context-Aware Federated Prompt Tuning},
  year={2024},
  volume={23},
  number={12},
  pages={14044-14054}
}

@InProceedings{sun2023fedperfix,
    author={Sun, Guangyu and Mendieta, Matias and Luo, Jun and Wu, Shandong and Chen, Chen},
    title={FedPerfix: Towards Partial Model Personalization of Vision Transformers in Federated Learning},
    booktitle={Proceedings of the IEEE/CVF International Conference on Computer Vision},
    month={October},
    year={2023},
    pages={4988-4998}
}

@inproceedings{bossard14food,
  title = {Food-101 -- Mining Discriminative Components with Random Forests},
  author={Bossard, Lukas and Guillaumin, Matthieu and Van Gool, Luc},
  booktitle={European conference on computer vision},
  pages={446--461},
  year={2014}
}

@article{tzu2019measuring,
  author = {Tzu{-}Ming Harry Hsu and Hang Qi and Matthew Brown},
  title  = {Measuring the Effects of Non-Identical Data Distribution for Federated Visual Classification},
  year = {2019},
  journal = {arXiv preprint arXiv:1909.06335}
}

@article{touvron2023llama,
  title={Llama: Open and efficient foundation language models},
  author={Touvron, Hugo and Lavril, Thibaut and Izacard, Gautier and Martinet, Xavier and Lachaux, Marie-Anne and Lacroix, Timoth{\'e}e and Rozi{\`e}re, Baptiste and Goyal, Naman and Hambro, Eric and Azhar, Faisal and others},
  journal={arXiv preprint arXiv:2302.13971},
  year={2023}
}

@inproceedings{peng2025look,
  title={Look Back for More: Harnessing Historical Sequential Updates for Personalized Federated Adapter Tuning},
  author={Peng, Danni and Wang, Yuan and Fu, Huazhu and Jiang, Jinpeng and Liu, Yong and Goh, Rick Siow Mong and Wei, Qingsong},
  booktitle={Proceedings of the AAAI Conference on Artificial Intelligence},
  volume={39},
  number={19},
  pages={19857--19865},
  year={2025}
}

@article{pfeiffer2024efficient,
  title={Efficient Federated Finetuning of Tiny Transformers with Resource-Constrained Devices},
  author={Pfeiffer, Kilian and Ahmed, Mohamed Aboelenien and Khalili, Ramin and Henkel, J{\"o}rg},
  journal={arXiv preprint arXiv:2411.07826},
  year={2024}
}

@ARTICLE{wang2025federated,
  author={Wang, Zixin and Zhou, Yong and Shi, Yuanming and Letaief, Khaled B.},
  journal={IEEE Transactions on Wireless Communications},
  title={Federated Fine-Tuning for Pre-Trained Foundation Models Over Wireless Networks},
  year={2025},
  volume={24},
  number={4},
  pages={3450-3464}
}

@article{li2004Learning,
  title={Learning Generative Visual Models from Few Training Examples: An Incremental Bayesian Approach Tested on 101 Object Categories},
  author={Fei-Fei, Li and Fergus, Rob and Perona, Pietro},
  journal={Computer Vision and Pattern Recognition Workshop},
  year={2004},
}

@inproceedings{nguyen2022where,
  title={Where to begin? exploring the impact of pre-training and initialization in federated learning},
  author={Nguyen, John and Malik, Kshitiz and Sanjabi, Maziar and Rabbat, Michael},
  journal={International Conference on Learning Representations},
  year={2023}
}

@inproceedings{chen2023on,
  title={On the Importance and Applicability of Pre-Training for Federated Learning},
  author={Chen, Hong-You and Tu, Cheng-Hao and Li, Ziwei and Shen, Han Wei and Chao, Wei-Lun},
  journal={International Conference on Learning Representations},
  year={2023}
}

@article{karimireddy2021breaking,
  title={Breaking the centralized barrier for cross-device federated learning},
  author={Karimireddy, Sai Praneeth and Jaggi, Martin and Kale, Satyen and Mohri, Mehryar and Reddi, Sashank and Stich, Sebastian U and Suresh, Ananda Theertha},
  journal={Advances in Neural Information Processing Systems},
  volume={34},
  pages={28663--28676},
  year={2021}
}

@inproceedings{liu2021Swin,
  title={Swin Transformer: Hierarchical Vision Transformer using Shifted Windows},
  author={Liu, Ze and Lin, Yutong and Cao, Yue and Hu, Han and Wei, Yixuan and Zhang, Zheng and Lin, Stephen and Guo, Baining},
  booktitle={Proceedings of the IEEE/CVF International Conference on Computer Vision (ICCV)},
  year={2021}
}

@inproceedings{sun2025exploring,
  author={Sun, Yuchang and Xie, Yuexiang and Ding, Bolin and Li, Yaliang and Zhang, Jun},
  title={Exploring Selective Layer Fine-Tuning in Federated Learning},
  booktitle={IEEE International Symposium on Information Theory (ISIT)},
  month={June},
  year={2025},
}

@article{li2026fedquad,
  author={Liu, Jianchun and Li, Rukuo and Xu, Hongli and Ma, Qianpiao and Yan, Jiaming and Huang, Liusheng},
  journal={IEEE Transactions on Mobile Computing},
  title={FedQuad: Adaptive Layer-Wise LoRA Deployment and Activation Quantization for Federated Fine-Tuning},
  year={2026},
  volume={25},
  number={5},
  pages={6320-6334}
}

@article{chen2024heterogeneity,
  title={Heterogeneity-guided client sampling: Towards fast and efficient non-iid federated learning},
  author={Chen, Huancheng and Vikalo, Haris},
  journal={Advances in Neural Information Processing Systems},
  volume={37},
  pages={65525--65561},
  year={2024}
}

@article{shi2026dystop,
  title={DySTop: Dynamic Staleness Control and Topology Construction for Asynchronous Decentralized Federated Learning},
  author={Shi, Yizhou and Ma, Qianpiao and Xu, Yan and Zhou, Junlong and Hu, Ming and Liao, Yunming and Xu, Hongli},
  journal={IEEE Transactions on Mobile Computing},
  year={2026},
  volume={25},
  number={8},
  pages={11662--11678},
  publisher={IEEE}
}

@inproceedings{zhang2024when,
    title={When Scaling Meets {LLM} Finetuning: The Effect of Data, Model and Finetuning Method},
    author={Biao Zhang and Zhongtao Liu and Colin Cherry and Orhan Firat},
    booktitle={The Twelfth International Conference on Learning Representations},
    year={2024}
}

@misc{kaplan2020scaling,
      title={Scaling Laws for Neural Language Models},
      author={Jared Kaplan and Sam McCandlish and Tom Henighan and Tom B. Brown and Benjamin Chess and Rewon Child and Scott Gray and Alec Radford and Jeffrey Wu and Dario Amodei},
      year={2020},
      eprint={2001.08361},
      archivePrefix={arXiv}
}

@inproceedings{fan2024scaling,
  title={Scaling laws of synthetic images for model training... for now},
  author={Fan, Lijie and Chen, Kaifeng and Krishnan, Dilip and Katabi, Dina and Isola, Phillip and Tian, Yonglong},
  booktitle={Proceedings of the IEEE/CVF Conference on Computer Vision and Pattern Recognition},
  pages={7382--7392},
  year={2024}
}

@inproceedings{kang2025demystifying,
    title = {Demystifying Synthetic Data in {LLM} Pre-training: A Systematic Study of Scaling Laws, Benefits, and Pitfalls},
    author = {Kang, Feiyang and Ardalani, Newsha and Kuchnik, Michael and Emad, Youssef and Elhoushi, Mostafa and Sengupta, Shubhabrata and Li, Shang-Wen and Raghavendra, Ramya and Jia, Ruoxi and Wu, Carole-Jean},
    booktitle = {Proceedings of the 2025 Conference on Empirical Methods in Natural Language Processing},
    month = {nov},
    year = {2025},
    publisher = {Association for Computational Linguistics},
    pages = {10739--10758}
}

@inproceedings{pearce2025scaling,
    title={Scaling Laws for Pre-training Agents and World Models},
    author={Tim Pearce and Tabish Rashid and David Bignell and Raluca Georgescu and Sam Devlin and Katja Hofmann},
    booktitle={Forty-second International Conference on Machine Learning},
    year={2025}
}

\setcounter{mylemma}{0}
\setcounter{mydefinition}{0}

\end{document}